# Spatially tailored spin wave excitation for spurious-free, low-loss magnetostatic wave filters with ultra-wide frequency tunability

Shuxian Wu[1,†], Shun Yao[1,†], Xingyu Du[1], Chin-Yu Chang[1], Roy H. Olsson III[1,*]

[1]Department of Electrical and Systems Engineering, University of Pennsylvania, Philadelphia, PA, USA.

[†]These authors contributed equally to this work.

[*]Author to whom correspondence should be addressed: rolsson@seas.upenn.edu

## ABSTRACT

Yttrium iron garnet magnetostatic wave (MSW) radio frequency (RF) cavity filters are promising for sixth-generation (6G) communication systems due to their wide frequency tunability. However, the presence of severe spurious modes arising from the finite cavity dimensions severely degrades the filter performance. We present a half-cone transducer that spatially tailors spin wave excitation to selectively enhance the primary cavity modes comprising the MSW filter passband, while strongly suppressing the undesired spurious modes. Theoretical analysis, numerical simulations and experiments verify the effectiveness of the spatially tailored technique. We utilize the half-cone transducer to demonstrate a spurious-free, single-cavity half-cone MSW filter (HC-MSWF) with an insertion loss (IL) of 2.4-3.2 dB over a frequency tuning range of 6.3-16.8 GHz. Extending our study, we further demonstrate a spurious-free, dual-cavity HC-MSWF with an unprecedented tuning range of 21.7 GHz (9.8-31.5 GHz) while maintaining a low IL of 2.9-3.8 dB. This significant advance in performance will enable highly reconfigurable and robust 6G networks.

## INTRODUCTION

State-of-the-art reconfigurable radios, such as cellular handsets, rely on a switched bank of more than 50 filters to operate across multiple bands while suppressing interference (*1*). With the rapid evolution and continuous expansion of wireless communication technologies, especially with the emerging sixth-generation (6G) networks, significantly more fixed frequency radio-frequency (RF) filters are required due to the continuous increase in the number of frequency bands in communication systems (*2*). Tunable RF filters have drawn widespread attention as an alternative to switched banks of fixed frequency filters due to their multi-frequency compatibility across different bands (*3*). Specifically, advanced software defined radio systems (*4*), radar technologies (*5*), base stations (*6*), satellite communications (*7*), and Internet of Things (IoT) devices (*8*) demand tunable filters with low insertion loss, spurious-free responses and high selectivity to ensure reliable and efficient spectrum operations.

Tunable RF filters dynamically select operational frequencies by employing various tuning mechanisms, thus significantly enhancing system flexibility and spectral efficiency. Varactor-based tunable filters typically leverage the voltage-dependent capacitance of semiconductor devices such as varactor diodes, PIN diodes, or field-effect transistors (FETs) to alter the resonant frequency of the filters. Tunable filters based on varactors offer advantages such as small size, low power consumption and fast tuning speed (*9–11*). However, their frequency tuning ratio is generally limited below 2:1, and the insertion loss is relatively high due to the low resonator quality factor (*Q*). Low-*Q* and high insertion loss are particularly problematic at millimeter-wave and sub-millimeter-wave frequency bands, restricting their applicability in advanced applications. Microelectromechanical systems (MEMS) technology enables the fabrication of miniature mechanical structures that can be actuated to change the physical dimensions or shapes of resonators, thus achieving frequency tuning (*12-15*). MEMS-based tunable filters exhibit potential for small size, high integration and low power consumption, but challenges remain in realizing high tuning speed and wide frequency tuning ratio of more than 2.5:1. Surface acoustic wave (SAW) and bulk acoustic wave (BAW) filters are widely used in RF front-ends due to their compact dimensions, low fabrication costs and high *Q* value (*16*, *17*). Frequency tuning can be achieved by incorporating variable capacitor or switching elements onto SAW/BAW resonators (*18*, *19*). However, SAW/BAW filters typically experience significant performance degradation at high frequency bands, and their frequency tuning range is extremely narrow being only a few percent of the center frequency.

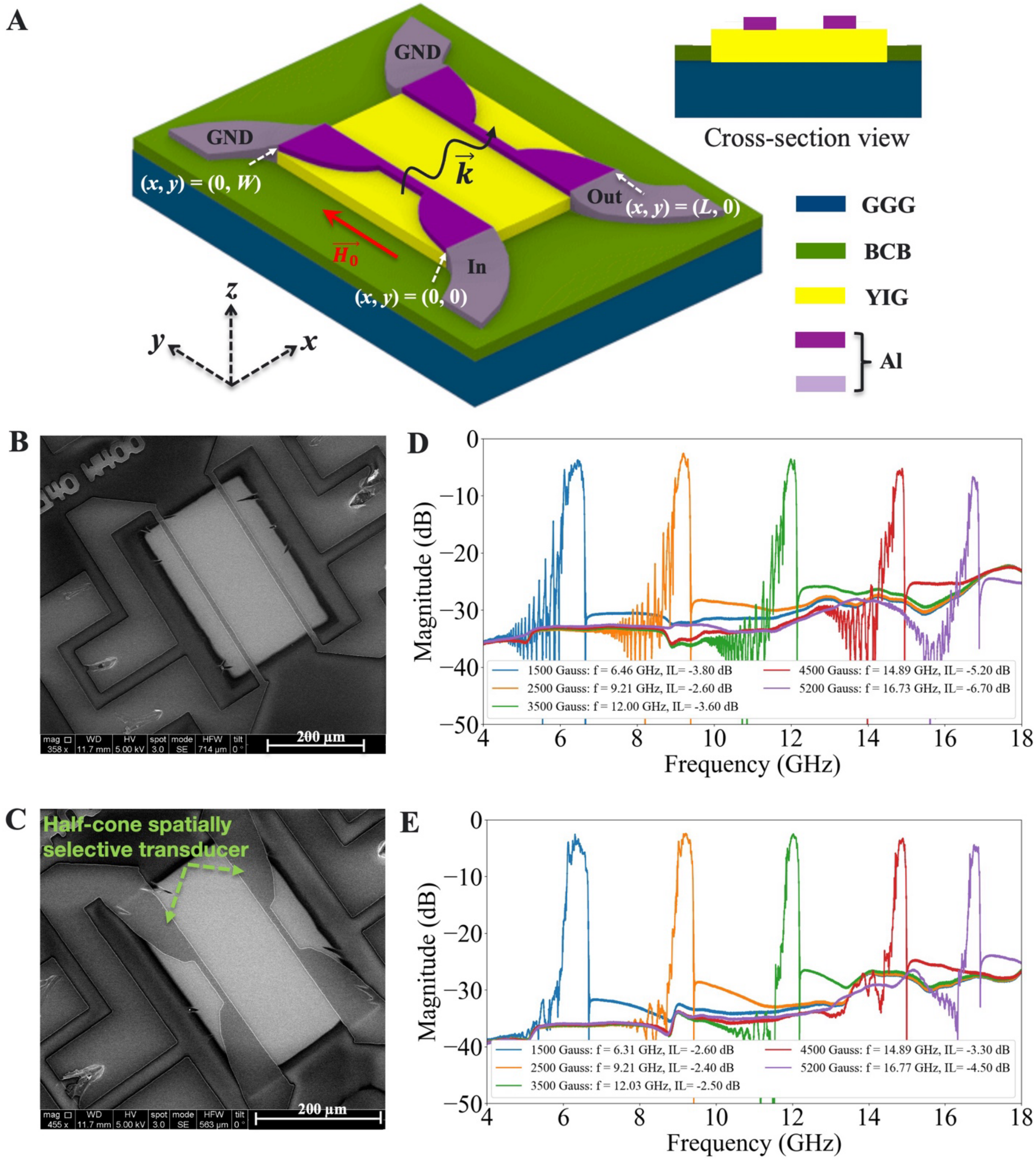

**Fig. 1. Design and realization of half-cone magnetostatic wave filters (HC-MSWF).** (**A**) 3-D geometry and cross section of the proposed HC-MSWF. A 15 μm thick YIG film grown on top of a Gadolinium Gallium Garnet (GGG) substrate is patterned by wet etching using phosphoric acid, and a negative photoresist Benzocyclobutene (BCB) is laid on top of the YIG for planarization. The Al transducers are patterned by dry etching for the conversion between electromagnetic waves and magnetostatic waves. An external bias magnetic field ($\vec{H}_0$) is applied to the YIG thin film. (**B**) SEM image of the fabricated single-cavity reference straight-line bandpass filter. (**C**) SEM image of the fabricated single-cavity bandpass filter with half-cone spatially selective transducer. (**D**) The measured frequency responses of the reference bandpass filter using straight-line transducers. It can be observed that there are many spurious modes on the left side of the filter passband, and the insertion loss increases significantly at higher frequency bands. (**E**) The measured frequency responses of the HC-MSWF. All spurious modes in the HC-MSWF are effectively suppressed and the insertion losses at all frequency bands are lower than that of the reference filters.

Research on the generation and propagation of spin waves in ferromagnetic films has promoted the development of spin-based device technology (*20*). Tunable filters based on spin waves using ferrite materials such as yttrium iron garnet (YIG) exhibit low insertion loss, high *Q*, and extremely large frequency tuning ranges from several GHz to tens of GHz when subjected to tunable external bias magnetic fields (*21–24*). These unique properties enable YIG-based tunable filters to provide outstanding performance, especially at microwave and millimeter-wave frequencies, making them highly attractive for next-generation communication systems. Commercially available YIG filters typically rely on the manual alignment of coupling transducer loops around YIG spheres, which significantly hinders scalability (*25*). Additionally, the traditional design results in large volume and high power-consumption due to the substantial currents required by the electromagnets used to tune the magnetic bias field and filter frequency. Recently, single-crystal thin-film YIG resonators and filters with small magnetic damping have been demonstrated, resulting in lower filter insertion loss and smaller size when compared with YIG sphere-based filters (*26–33*). An edge-coupled magnetostatic forward volume wave bandpass filter (MSFVW) was demonstrated with an insertion loss of 6.94 dB from 4.5 GHz to 10.1 GHz (*28*). A microfabrication deep-etching methodology was studied to enlarge the coupling of thin film YIG magnetostatic wave resonators (*29*). To solve the problem of high power consumption, a YIG film-based magnetostatic surface wave (MSSW) filter was proposed with relatively wide frequency tuning range from 3.4 GHz to 11.1 GHz, insertion loss of 3.2–5.1 dB, and zero static power consumption by developing a novel tunable magnetic bias circuit (*31*). YIG tunable filters have demonstrated frequency tunability from 4–18 GHz along with nonreciprocal behavior (*32*). In addition, thick YIG-based two-cavity tunable filters have achieved wide tunability with insertion loss below 5 dB (*33*).

However, a significant challenge in the advancement of YIG-based MSSW filters is the presence of severe spurious responses. The unwanted modes in MSSW filters primarily arise from several sources: a) For finite width YIG films, magnetostatic waves not only propagate along the length direction of YIG (*x*-direction in Figure 1A) but also form standing wave modes across the width direction of the film (*y*-direction in Figure 1A), which may generate unwanted width modes (*34*). b) The strong dispersion of magnetostatic waves in YIG-based bandpass filters can lead to the unintended excitation of higher-order modes due to the frequency-dependent variation in phase and group velocities, resulting in the excitation of spurious responses (*35*). c) Reflections of MSSW at the input/output ports and the edges of the film may interfere, causing some spurious responses (*36*). Such spurious responses degrade filter selectivity and spectral purity, severely impacting their practical applicability in advanced RF systems.

Researchers have explored various techniques to mitigate the adverse effects of spurious modes in magnetostatic spin wave (MSW) filters. High-quality YIG films grown by liquid phase epitaxy can reduce intrinsic losses and spurious modes generation (*37*). Designing specialized input/output coupling structures, such as multi-section couplers or rotated YIG slabs with a microstrip transducer pair, can suppress the excitation of spurious modes (*38*, *39*). Although these techniques have demonstrated success in reducing spurious modes, they often complicate filter design, fabrication procedures and inadvertently increase insertion loss, thus presenting challenging trade-offs, particularly across extremely wide tuning ranges. Previous studies exploited the unique dispersion characteristics of 15–18 μm thick YIG thin films and demonstrated partial attenuation of spurious modes at high frequency bands, achieving 28 dB isolation at frequencies 0.9 times the bandwidth above the center frequency (*32*, *33*). However, spurious suppression over a wide tunable range remains a critical unresolved challenge. Moreover, there is a research gap in numerical analysis of the MSSW modes and the corresponding spurious suppression techniques for micromachined thin-film YIG filters to date. Effective mitigation of these spurious responses is crucial for realizing the full potential of thin film YIG-based tunable filters.

Apodization, also known as aperture or electrode weighting, is a technique employed to optimize the frequency response or spatial radiation characteristics of transducers in acoustic applications such as SAW and BAW filters (*40–42*). Through the deliberate design of the electrodes, the amplitude/phase or acoustic velocity distribution of excitation waves along transducers is modified to guide the coupling of waves. This technique effectively suppresses the spurious responses of acoustic filters, significantly enhancing the selectivity without notably increasing insertion loss or manufacturing complexity (*43–46*). The proven effect of apodization techniques in acoustic filters provides strong motivation for addressing the similar spurious suppression challenges in MSSW filters.

In this study, inspired by the successful application of apodization technique in acoustics, we propose a novel spatially tailored half-cone transducer for MSSW bandpass filters, demonstrating suppression of spurious modes over an unprecedented frequency tuning range of 6.3–31.5 GHz. Figure 1A depicts a 3-D schematic geometry and a cross-section of the half-cone magnetostatic wave filter (HC-MSWF), where *L* represents the length of cavity in the direction of MSSW propagation (*x*-direction), and *W* represents the width of cavity along the length direction of the transducers (*y*-direction). A negative photoresist Benzocyclobutene (BCB) is laid on top of YIG for planarization. To comprehensively analyze and visualize the mode shapes in thin-film YIG filters, finite element method (FEM) simulations are conducted in Ansys High Frequency Structure Simulator (HFSS) software platform, providing a powerful guide for the spatially selective transducer technique. Combining with Maxwell's equations and Landau-Lifshitz-Gilbert equation analysis, a half-cone transducer is introduced to tailor the excitation of spin waves. The current-density distribution of the half-cone transducers is modulated to excite the desired primary modes and to suppress the spurious modes. Meanwhile, the half-cone transducer can reduce in-band insertion loss due to a better impedance matching. Simulation results demonstrate that the half-cone transducer can effectively suppress undesired and strong width modes in conventional MSSW filters using straight-line (unapodized) transducers.

To further validate the feasibility and effectiveness of this novel approach, single-cavity and dual-cavity MSSW bandpass filters using the half cone spatially selective transducer are fabricated. Scanning electron microscope (SEM) images of the manufactured single-cavity reference straight-line and half-cone bandpass filters are observed in Figure 1 (B and C), where the two pads located on either side of the YIG cavity implement the RF input/output transducers. The measured frequency responses of single-cavity filters at different bias magnetic field intensities implemented with traditional straight-line transducers and spatially tailored half-cone transducers are depicted in Fig. 1D-E. Experimental results closely match the simulations, highlighting significant improvements in spurious mode suppression. The half-cone transducer technique successfully suppresses all spurious modes and reduces the insertion loss across multiple frequency bands. The insertion loss of filters in this study is 2.4–3.8 dB across the extremely broad frequency tuning range of 6.3–31.5 GHz, indicating superior performance compared to switched filter banks or alternative YIG thin film-based tunable filter technologies.

## RESULTS

### Magnetostatic surface wave cavity behavior

MSWs in ferrite media exhibit multiple distinct propagation characteristics depending on the orientation of an external bias magnetic field ($\vec{H}_0$) relative to the wave vector ($\vec{k}$) and the normal vector ($\vec{n}$) of the surface of the ferrite slabs. Damon and Eshbach originally categorized the MSWs into three distinct propagation modes including MSSW, magnetostatic forward volume waves (MSFVW) and magnetostatic backward volume waves (MSBVW), which are foundational in understanding MSW dynamics (*47*).

For a straight MSSW resonator based on thin-film YIG, when the bias magnetic field is perpendicular to the wave propagation direction, surface waves propagate along the surface of the YIG film and reflect at the straight boundaries onto the other surface (*48*). A circular wave pattern is established when the following resonance condition is satisfied (*49*),

$$2k_{xn}L = 2n\pi, \tag{1}$$

where $k_{xn}$ is the average wavenumber on the top and bottom surfaces, and *n* is an integer representing the order of the spin waves in the *x* direction (primary modes). As *L* increases, the spacing between the different primary modes gets smaller due to the strong dispersion of the MSSWs. When the thickness of the YIG thin film is much smaller than the width, a width mode can be excited in the *y*-direction, the wavenumber ($k_{ym}$) can be described as a simple expression:

$$k_{ym} = \frac{m\pi}{W}, \tag{2}$$

where *m* represents the order of spin waves in the *y* direction (width modes).

In YIG thin-film–based filters, the interaction between transducer-induced currents and magnetization dynamics is governed by the coupled Maxwell's equations and Landau–Lifshitz–Gilbert (LLG) equation (*50*). Current density (*J*) in patterned transducers generates an oscillating magnetic field ($H_{\mathrm{rf}}$), as described by Ampère's law in Maxwell's equations:

$$\nabla \times H_{\mathrm{rf}}(r,t) = J(r,t) + \frac{\partial D}{\partial t}, \tag{3}$$

where *r* is the spatial domain variable, *t* is the time domain variable, J is the electric current density, and *D* is the electric displacement. This field penetrates the YIG film and acts as the primary excitation source for magnetostatic spin waves. The temporal evolution of the local magnetization (*M*) is then dictated by the LLG equation under the influence of an effective magnetic field ($H_{\mathrm{eff}}$):

$$\frac{\partial M(r,t)}{\partial t} = \mu_0 \gamma (M \times H_{\mathrm{eff}}) + \frac{\alpha}{|M|} M \times \frac{\partial M}{\partial t} + \tau, \tag{4}$$

$$H_{\mathrm{eff}} = H_{\mathrm{rf}} + H_0 + H_{\mathrm{ani}} + H_{\mathrm{exch}}, \tag{5}$$

where $\mu_0$ is the vacuum permeability, $\gamma$ represents the gyromagnetic ratio, $\alpha$ is the Gilbert damping constant, and $\tau$ represents a physical property unique to spintronic devices. The first term of LLG equation provides an expression for the frequency dependence of the magnetizing response. Combining the expression of magnetic flux density ($B = \mu_0(H$

$+ M$)), the component form of lossless permeability tensor ($\mu$) in the YIG film can be characterized as:

$$\mu = \begin{bmatrix} \mu_1 & \mu_2 & 0 \\ -i\mu_2 & \mu_1 & 0 \\ 0 & 0 & 1 \end{bmatrix}, \tag{6}$$

where $\mu_1 = 1 - \frac{\Omega_H}{\Omega^2 - \Omega_H^2}$, $\mu_2 = \frac{\Omega_H}{\Omega^2 - \Omega_H^2}$, $\Omega = \frac{\omega}{\gamma M_s}$, $\Omega_H = \frac{H_0}{M_s}$, ω is the angular frequency of the magnetostatic waves, $\gamma$ is the gyromagnetic ratio ($\gamma_{\text{YIG}} = 2.8 \times 10^6$ Hz/Gauss), and $M_s$ represents the saturation magnetization of YIG thin film. In free space, the magnetic permeability is described by a scalar quantity with a normalized value of unity; accordingly, as expressed in the notation of Eq. (1), this corresponds to $\mu_1 = 1$ and $\mu_2 = 0$.

In the condition of the magnetostatic approximation ($\nabla \times H_0 = 0$), the magnetic field can be expressed as the gradient of a scalar potential ($\psi$). After using the simplifying assumption, $H_0$ can be written as:

$$H_0(r) = -\nabla\psi(r), \tag{7}$$

where $r$ is the three-dimensional vector of positions. In non-magnetic materials, the magnetic flux density is expressed as:

$$B(r) = \mu \cdot H_0(r). \tag{8}$$

Requiring the divergence of B($r$) to vanish, together with Equations (6) – (7), results in the following equation:

$$\left[\mu_1\left(\frac{\partial^2}{\partial x^2} + \frac{\partial^2}{\partial y^2}\right) + \frac{\partial^2}{\partial z^2}\right]\psi(r) = 0, \tag{9}$$

where an appropriate value $\mu_1$ can be used inside and outside the YIG film. To approximate the boundary pinning conditions at the sample edges, we assume that the transverse components of the microwave magnetic field vanish, i.e., $B_y = 0$, at the boundaries of $y = 0$ and $y = W$ in Figure 1a. Under this assumption, a solution to Eq. (9) can be formulated within the region $0 \leq y \leq W$ as (*48*):

$$\psi(r) = X(x)e^{ik_{ym}y}\sin(k_z z), \tag{10}$$

After combining Equations (9) and (10), the wavenumber in the $z$-direction inside the YIG film can be expressed as:

$$k_z^2 = k_{xn}^2 + k_{ym}^2 = k_{xn}^2 + \frac{\left(\frac{m\pi}{W}\right)^2}{\mu_1}, \tag{11}$$

and outside

$$k_z^2 = k_{xn}^2 + k_{ym}^2 = k_{xn}^2 + \left(\frac{m\pi}{W}\right)^2. \tag{12}$$

In a first order approximation, the normal spin-wave modes of magnetic waveguides can be treated according to the dipole exchange theory of spin-wave spectra in extended magnetic films (*51*). The significant advantage of this theory is that it provides an explicit relation between the wave vector and the spin-wave frequency ($f$), which significantly simplifies the analysis. According to (*52*), the relation can be written as:

$$f^2 = \gamma^2[H_0 + 4\pi M_0(1 - P + \alpha k_z^2)] \times \left[H_0 + 4\pi M_0\left(P\frac{k_{ym}^2}{k_z^2} + \alpha k_z^2\right)\right], \tag{13}$$

where $P = 1 - (1 - \exp(-k_z T))/(k_z T)$, $\alpha$ is the exchange stiffness ($\alpha_{YIG} = 5.18 \times 10^{-13}$ cm$^2$), and $T$ is the film thickness of YIG. This expression can calculate the dispersions of MSSW and MSBVW. The dispersions of the two waves start at the minimum frequency of the uniform ferromagnetic resonance ($f_{min}$):

$$f_{min} = \sqrt{\gamma(H_0(H_0 + M_s))}, \tag{14}$$

and the maximum convergence frequency ($f_{max}$) of MSSW can be expressed as:

$$f_{max} = \gamma\sqrt{H_0(H_0 + M_s) + \frac{M_s^2}{4}}. \tag{15}$$

For the MSSW bandpass filters in this study, a 15 μm thick YIG thin-film is grown on top of a Gadolinium Gallium

Garnet (GGG) substrate. Two transducers are adopted to convert energy between electrical currents and magnetostatic waves. The external bias magnetic field parallel to the transducers is applied to the YIG thin film. In such a setup, MSSW will be excited and propagate in a direction perpendicular to the transducers. To clearly observe the frequency changes with wavenumber in the *x*-direction for the primary modes and width modes in YIG thin-film, the dispersion curves at a magnetic field intensity of 2500 Gauss are described in Part 1 of the supplementary materials.

**Magnetostatic wave cavity mode shapes**

In the study of acoustic filters, modal analysis helps guide the design and thus improves the performance of filters. However, there has been little work on modal analysis for MSW cavity devices to date. In this work, a three-dimensional (3-D) electro-magnetostatic wave simulation model is built for the cavity bandpass filters using a finite element method in the Ansys HFSS platform. The top view of the 3-D simulation model is depicted (Fig. 2A), where the resonant body including a GGG layer, a YIG thin film layer (yellow) and an Al layer (dark blue) are placed in the middle area. The BCB layer is not included in the model to reduce computational complexity, as the BCB layer has almost no effect on the magnetostatic wave propagation. The length and width of the YIG are set as 280 μm and 400 μm to maximize the propagation of the primary modes. The thickness of GGG substrate is 500 μm and the thickness of the YIG thin film is 15 μm. The thick YIG thin film is chosen due to the demonstrated advantages to minimize spin-wave propagation loss, enhance magnetic coupling efficiency and improve power handling capacity, owing to the increased interaction volume of magnetic material with the magnetic fields produced by the transducers and thermal robustness of thicker films (*32*, *33*). A 2 μm thick Al film with a width of 10 μm is located at the *L*/4 position of the YIG to optimize the coupling efficiency and out-of-band rejection capability of the filters.

Two wave ports and ground-signal-ground (GSG) pads are located on both sides of the resonant body for input/output RF electrical signals. Compared to lumped port simulation, wave port simulation is more appropriate to observe high wave propagation and need to be suppressed to enhance filter performance. To facilitate identification of different modes, we denote the primary modes as "p", the width modes as "w", and the subsequent numbers express the order of the modes. For instance, "p2w3" refers to the third-order width mode of the second-order primary mode. The magnetic field vector distributions of the nine resonance modes along the thickness direction (*z*-direction) of the YIG thin film are observed (Fig. 2C). It can be observed that the desired primary modes which form the passband response are predominantly concentrated in the central region in the *y*-direction, whereas the width modes forming the undesired spurious modes extend towards the two edges of the YIG film. Consequently, a spatially selective excitation technique by tailoring the shape of the transducers is proposed to enhance the primary mode coupling (w = 1) while suppressing generation of width modes (w > 1).

**Half-cone spatially selective MSW excitation**

In conjunction with Equations (3) – (5), the current density of transducers affects *M* indirectly by generating $H_{\mathrm{rf}}$, which precesses the magnetization vector and excites spin waves in YIG. The spatial distribution of current density directly determines the local excitation efficiency and modal selectivity of the spin waves. Therefore, a spatially tailored half-cone transducer is designed to modulate the current density distribution for selectively exciting the desired spin waves (w = 1) while suppressing spurious responses (w > 1).

Through the previous simulations, various propagation modes can be clearly observed. The magnetic field vector distribution of the fourth-order primary mode with first-order width mode (p4w1) in the reference cavity bandpass filter is illustrated (Fig. 3A). The propagation of magnetostatic waves predominantly concentrates in the middle region of the YIG thin film along the width direction. To suppress the width modes as mentioned earlier, three spatially tailored transducer techniques including a full-cone, an extended cone, and a half-cone are studied by tailoring the shapes of the transducers in this work. Here we focus on the introduction of full-cone and half-cone techniques, and the extended cone technique is introduced in the supplementary materials. By precisely adjusting the electrode width of the transducers, the current density distribution changes along the length of the transducers, consequently affecting the dynamic magnetization within the YIG film and the mode coupling characteristics. The detailed information about current density distribution is introduced in Part 2 of the supplementary materials. When the current density at the transducer edges is lower than that at the central region, the width modes generated across the width direction is weakened, thereby effectively suppressing the width spurious modes.

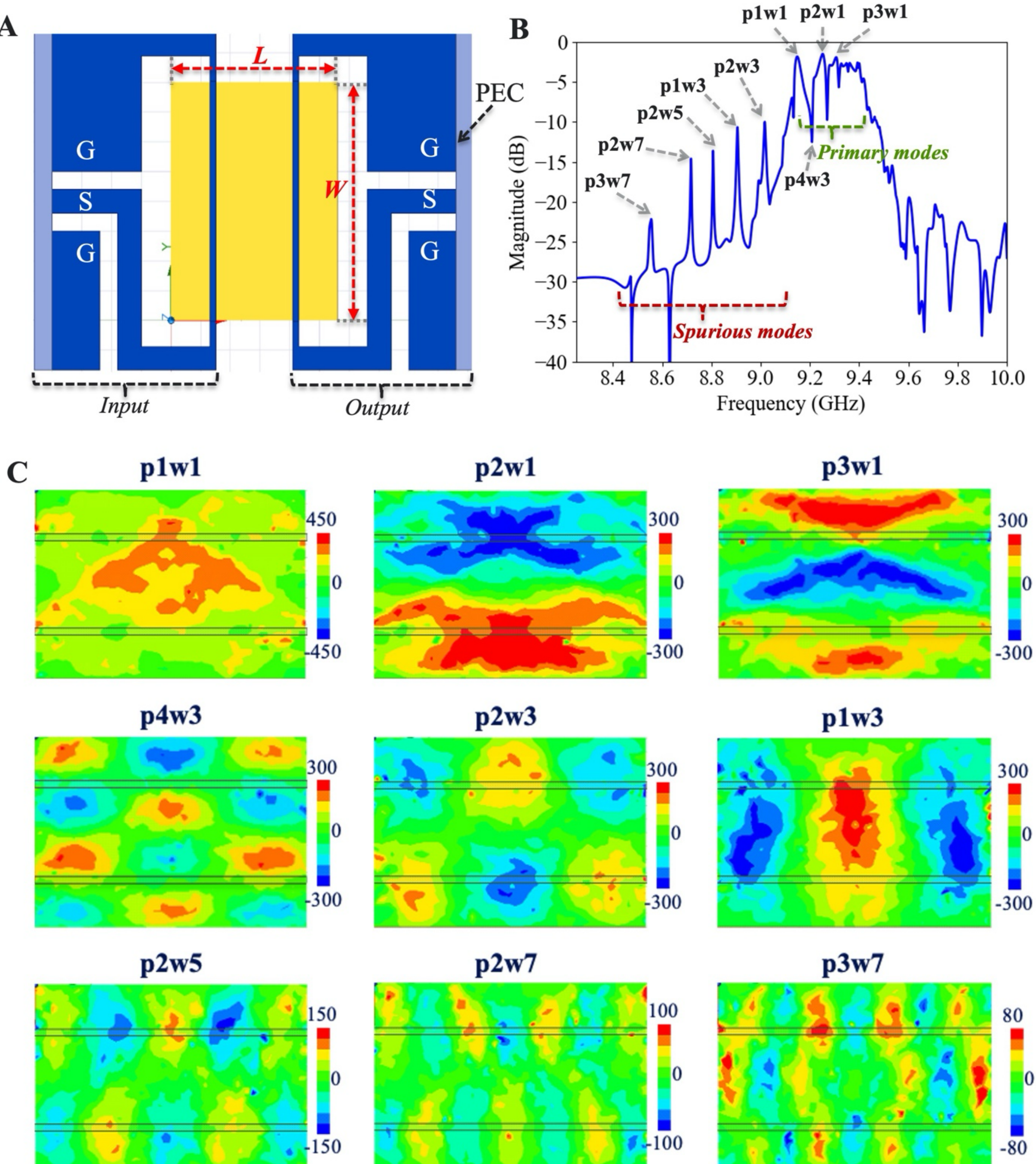

**Fig. 2. Simulations and magnetostatic wave mode shapes of the YIG thin film cavity MSWF. (A)** Top view of the simulation model of the MSWF cavity filters including a GGG layer, a YIG cavity layer (yellow) and an Al layer (dark blue). Two wave ports with perfect electric conductors (PEC) are applied on both sides of YIG for the radio-frequency signal input and output. **(B)** Simulated frequency response of the conventional straight-line transducer (reference) cavity bandpass filter. The different magnetostatic wave modes including primary modes and width modes are labeled in the frequency response, where "p" refers the order of the primary modes and "w" refers the order of the width modes. **(C)** Magnetic field vector shapes in the *z*-direction are shown for the primary and width magnetostatic wave modes with different orders in the reference cavity MSWF, where the units of all color bars are A/m.

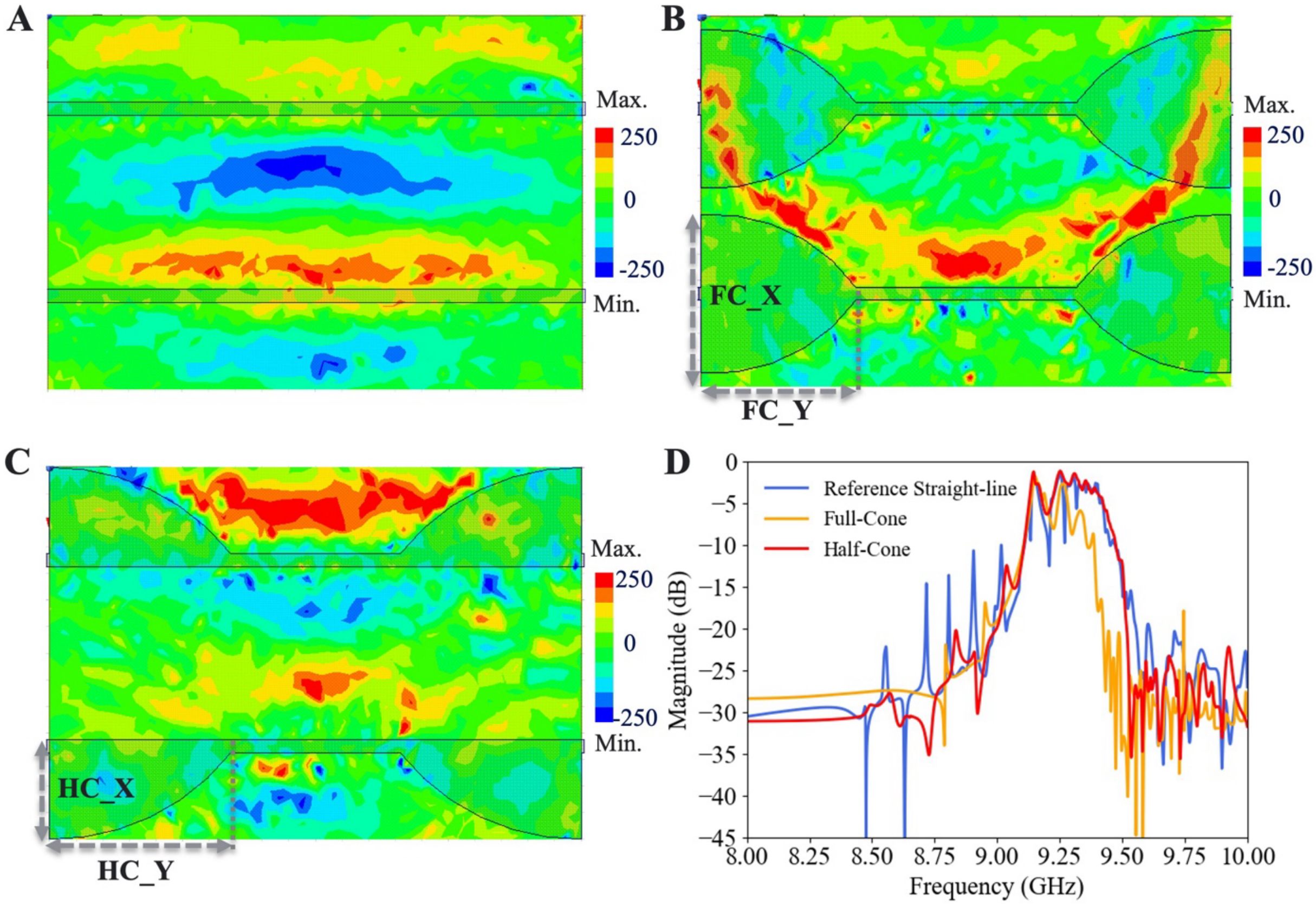

**Fig. 3. Comparison of the simulated magnetic field vector shapes for the high-order primary mode (p4w1) in the reference straight-line cavity filter, FC-MSWF and HC-MSWF, where the units of all color bars are A/m. (A)** Modes shape of the reference straight-line cavity bandpass filter. **(B)** Mode shape of the full-cone transducer cavity bandpass filter. **(C)** Mode shape of the half cone transducer cavity bandpass filter. **(D)** Simulated frequency responses of the three cavity bandpass filters. It can be observed that both the full-cone and half-cone transducers have a strong spurious suppression ability. The full-cone transducer increases the insertion loss and worsens the passband performance of filters, while the insertion loss of filters with the half-cone transducer matches the reference straight-line filters while suppressing the width spurious modes below the passband frequency.

First, a full-cone transdcuer is adopted at both edges of the two transducers. In this case variation of the electrode width causes the spacing between the two transducers to reduce, resulting in a difference of the MSSW phase at the receiver transducer. Consequently, high-order primary modes (e.g., p4w1 and p5w1) propagating from the edge of the full-cone region of one transducer toward another transducer are affected. The p4w1 mode shape of the full-cone MSWF is destroyed with weaker magnetostatic wave coupling compared to the reference straight-line filter (Fig. 3B), leading to increased insertion losses for the high-order primary modes that make up the upper frequency region of the filter passband. Fortunately, the half-cone transducer effectively addresses this issue (Fig. 3C). Due to the consistent spacing between the two transducers, the half-cone transducer not only suppresses the generation of width modes but also preserves the propagation characteristics of the higher-order primary modes. To compare the effects of the three transducers, the simulated frequency responses of the reference straight-line filter, full-cone MSWF (FC-MSWF) and HC-MSWF are presented (Fig. 3D). As anticipated, the full-cone transducer suppresses width spurious modes but introduces a higher insertion loss at the high end of the passband. While the half-cone transducer achieves effective spurious mode suppression and reduces the insertion loss simultaneously, thereby improving the overall filter performance. To further achieve a better spurious suppression and obtain a flatter passband response, the dimensions of the cavity HC-MSWF are optimized. The detailed information about HC-MSWF optimization is shown in Part 3 of the supplementary materials.

**Experimental results**

A scanning electron microscope (SEM) image of the fabricated single-cavity bandpass filter featuring half-cone transducers is shown in Fig. 1C. The white region indicates the YIG thin film cavity surrounded by a BCB layer, and the Al layer is put on the top of the YIG layer successfully. The fabricated devices are measured on a magnetic probe station using a vector network analyzer (VNA), which is described in detail in the Part 4 of the supplementary materials. The measured $S_{12}$ frequency responses of two bandpass cavity filters are analyzed, where the blue traces correspond to the reference straight-line bandpass cavity filter and the red traces represent the half-cone bandpass cavity filter (Fig. 4C-D). It can be observed that the half-cone transducer design effectively suppresses all width spurious modes ($w > 1$) of the cavity filter. The suppression level of spurious cavity modes is up to 15 dB. A comparison of the insertion loss (IL) for the two bandpass cavity filters are depicted (Fig. 4E). The insertion loss of the HC-MSWF (2.4–3.2 dB) is lower than that of the reference bandpass filter (2.6–5.2 dB), exhibiting a maximum reduction in IL of about 2 dB. The reduction of insertion loss arises from the decrease of equivalent electrode impedance ($Z_s$) in the filter with the half-cone geometry, leading to improved impedance matching for the filter input/output ports, which is especially pronounced at high frequencies. Additionally, bandpass filters employing the full-cone and extended-cone transducers discussed before were also fabricated and analyzed in this work, and the comparison of the measured results with the reference straight-line cavity filter are shown in Part 5 of the supplementary materials. By comparing the advanced techniques, it is concluded that the half-cone transducer technique exhibits superior performance in spurious mode suppression and effectively reduces the insertion loss of the bandpass filters, enabling exceptional performance over an extremely wide tuning range.

In the single-cavity reference straight-line bandpass cavity filters, the $Z_s$ is best matched with standard 50 Ω input/output port at impedances around 9 GHz. However, as the filter is tuned to higher frequency bands, the $Z_s$ increases accordingly, leading to impedance mismatch. Such mismatches inevitably result in increased insertion loss for the bandpass filters. To achieve low insertion loss bandpass filters at higher frequency bands, a reference straight

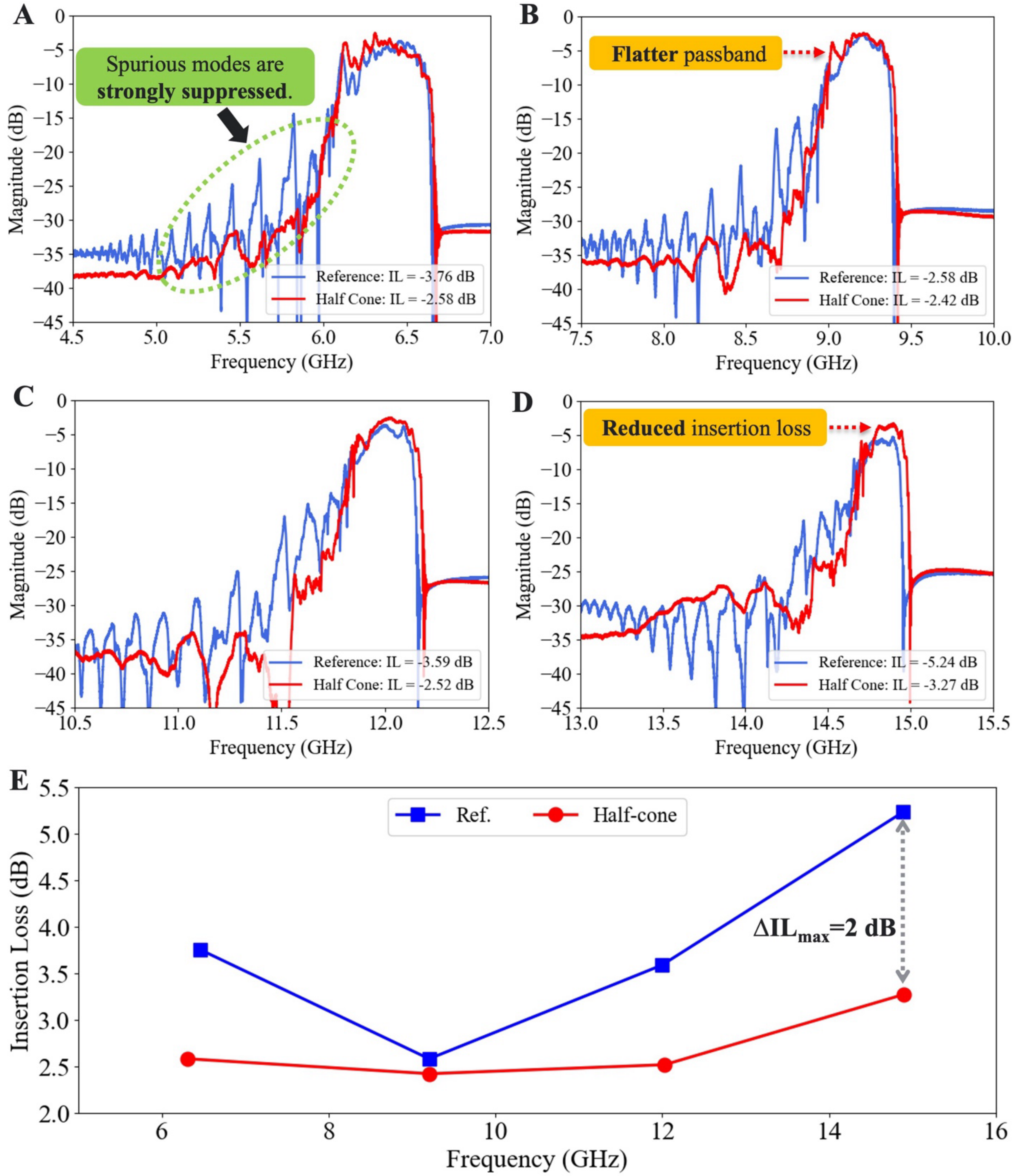

**Fig. 4. Measured frequency responses ($S_{12}$) of the reference straight-line and the half-cone bandpass filters at different magnetic field intensities. (A)** 1500 Gauss. **(B)** 2500 Gauss. **(C)** 3500 Gauss. **(D)** 4500 Gauss. **(E)** Insertion losses of the reference straight-line and half-cone cavity bandpass filters. The half-cone transducer effectively suppresses the width modes of the filters, and the insertion loss of the half-cone-based single-cavity bandpass filter is lower than that of the reference bandpass filter due to better impedance matching, especially at high frequency bands.

line and a half-cone dual-cavity bandpass filter are designed and fabricated. SEM images of the filters are observed, where two YIG cavities are connected in parallel (Fig. 5A-B). The design parameters of the dual-cavity filter are

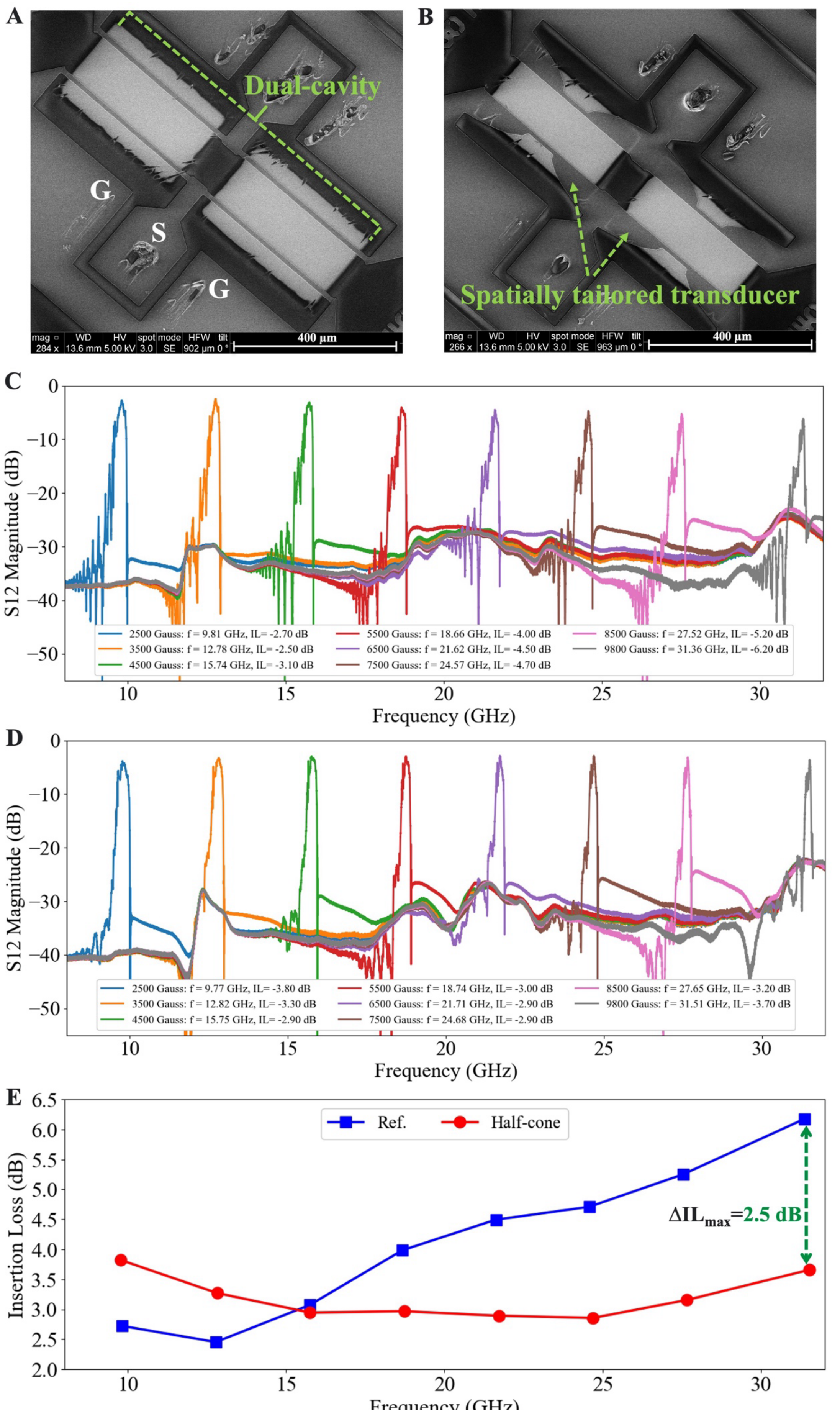

A
Dual-cavity
G
S
G
mag □ WD HV spot mode HFW tilt
284 x 13.6 mm 5.00 kV 3.0 SE 902 μm 0 °
400 μm
B
Spatially tailored transducer
mag □ WD HV spot mode HFW tilt
266 x 13.6 mm 5.00 kV 3.0 SE 963 μm 0 °
400 μm
C
S12 Magnitude (dB)
Frequency (GHz)
2500 Gauss: f = 9.81 GHz, IL= -2.70 dB
3500 Gauss: f = 12.78 GHz, IL= -2.50 dB
4500 Gauss: f = 15.74 GHz, IL= -3.10 dB
5500 Gauss: f = 18.66 GHz, IL= -4.00 dB
6500 Gauss: f = 21.62 GHz, IL= -4.50 dB
7500 Gauss: f = 24.57 GHz, IL= -4.70 dB
8500 Gauss: f = 27.52 GHz, IL= -5.20 dB
9800 Gauss: f = 31.36 GHz, IL= -6.20 dB
D
S12 Magnitude (dB)
Frequency (GHz)
2500 Gauss: f = 9.77 GHz, IL= -3.80 dB
3500 Gauss: f = 12.82 GHz, IL= -3.30 dB
4500 Gauss: f = 15.75 GHz, IL= -2.90 dB
5500 Gauss: f = 18.74 GHz, IL= -3.00 dB
6500 Gauss: f = 21.71 GHz, IL= -2.90 dB
7500 Gauss: f = 24.68 GHz, IL= -2.90 dB
8500 Gauss: f = 27.65 GHz, IL= -3.20 dB
9800 Gauss: f = 31.51 GHz, IL= -3.70 dB
E
Insertion Loss (dB)
Frequency (GHz)
Ref.
Half-cone
ΔIL_max=2.5 dB

**Fig. 5. Fabricated dual-cavity bandpass filters and measured results. (A)** SEM image of the reference straight-line dual-cavity bandpass filter. **(B)** SEM image of the spatially tailored half-cone dual-cavity bandpass filter. **(C)** The measured $S_{12}$ frequency responses of the reference dual-cavity bandpass filters. **(D)** The measured $S_{12}$ frequency responses of the half-cone dual-cavity bandpass filters. The half-cone transducer technique effectively suppresses spurious responses at all the frequency bands. **(E)** Insertion loss comparison of the reference and half-cone dual-cavity bandpass filters. The insertion loss of the half-cone dual-cavity filter above 15.75 GHz is lower than that of the reference dual-cavity straight-line filter. Using the spatially tailored transducer technique, a spurious-free and low insertion loss bandpass filter is achieved with a wide center frequency tuning range from 9.8 GHz to 31.5 GHz.

identical to those of a single cavity filter except for the number of cavities. By connecting two cavities in parallel, the total $Z_s$ of the filter is effectively halved and the impedance matching is improved, thereby reducing the IL especially at high frequencies. Moreover, the half-cone spatially tailored transducer is employed to suppress the spurious responses and to further reduce IL of the dual-cavity bandpass filters.

The measured frequency responses of the dual-cavity reference straight-line filters and HC-MSWF under different bias magnetic field intensities are illustrated (Fig. 5C-D). It is evident that the half-cone transducer effectively suppresses spurious responses across all the frequency bands, significantly enhancing the selectivity and improving the performance of filters. The out-of-rejection of the dual-cavity HC-MSWF is up to 41 dB at 9.8 GHz. The minimum insertion losses of the two filters over the tunable frequency range from 9.8 GHz to 31.5 GHz are compared in Fig. 5E. For frequency bands above 15.75 GHz, the insertion loss of the HC-MSWF is much lower than that of the reference straight line bandpass filter. The maximum difference in insertion loss between the two dual-cavity bandpass filters is 2.5 dB. The improved performance for the half-cone design at high frequency bands is attributed to superior impedance matching between the $Z_s$ and the port impedance. For clarity, Smith charts of the measured $S_{11}$ for the reference straight-line and the HC transducer dual-cavity bandpass filters are depicted and analyzed in Part 6 of the supplementary materials.

A performance comparison of the tunable bandpass filters developed in this study and other works is presented in Figure 6A-D, including single-cavity and dual-cavity configurations compared with previously reported varactor, MEMS, MEMS capacitor, and YIG thin film-based technologies (*11*), (*14*), (*28*), (*31*), (*53–58*). The half-cone transducer bandpass filter has a much larger frequency tuning range and smaller dimensions than other works based on varactor, MEMS, and MEMS capacitors. Moreover, the bandwidth of this work is larger, the frequency tuning range is much larger, and the out-of-band rejection is higher, especially in the higher frequency bands, when compared to other YIG thin film-based tunable passband filters. The single-cavity HC-MSWF in this work demonstrates a wide frequency tuning range of 6.3–16.8 GHz, a low insertion losses between 2.4 dB and 3.2 dB and a relatively large 3 dB bandwidth of approximately 245 MHz. Meanwhile, the dual-cavity configuration extends the large frequency tuning capability to 9.8–31.5 GHz with low insertion losses of 2.9–4.6 dB. The out-of-band rejection performance is comparable to that of most existing YIG thin film-based filters in the low-frequency range, and superior in the high-frequency range. To compare the level of spurious responses, the measured $S_{12}$ of the single-cavity HC-MSWF and other reported YIG film-based cavity filters are plotted together in Figure 6E-F. The half-cone transducer technique in this study effectively suppresses the spurious modes of the filters and achieves a relatively larger bandwidth, which outperforms existing state-of-the-art works in terms of performance.

The experimental results underscore the effectiveness of the half-cone spatially tailored transducer for spurious-free, low insertion loss, wide frequency tuning range and comparable out-of-band rejection characteristics, positioning the tunable filters described herein as highly suitable candidates for next-generation wireless and RF communication applications. The detailed performance comparison of this study and other works is listed in Table S1 in Part 7 of the supplementary materials.

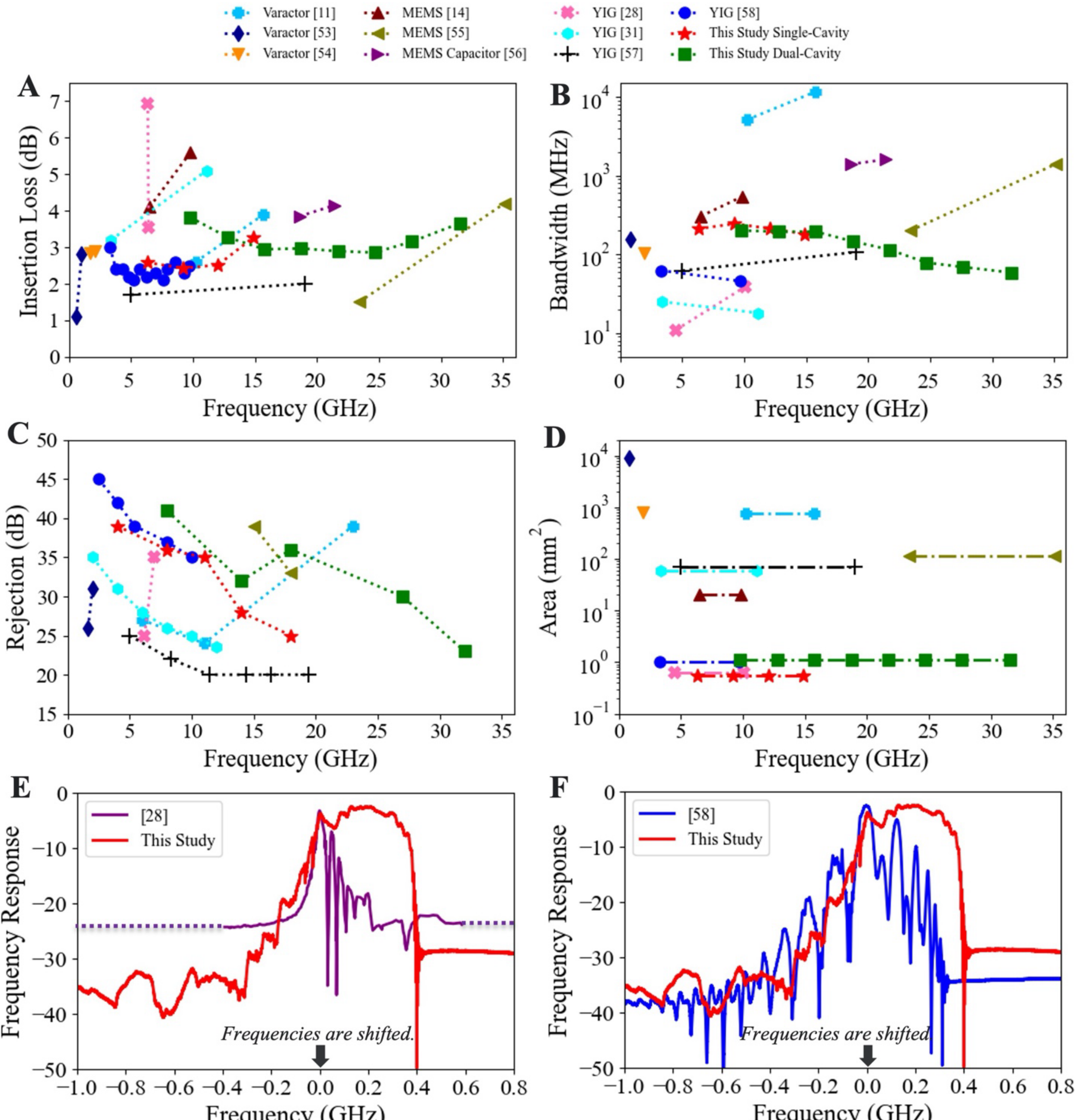

**Fig. 6. Performance comparison of varactor, MEMS, MEMS capacitor, and YIG thin film-based tunable bandpass filters.** **(A)** Insertion loss. **(B)** 3-dB bandwidth. **(C)** Out-of-band rejection. **(D)** Area of the fabricated tunable filters. Frequency response of the single-cavity HC-MSWF at a magnetic field intensity of 2500 Gauss compared with **(E)** Ref. (*28*) and **(F)** Ref. (*58*). The HC-MSWF has a much larger tuning range and a smaller dimension than other works based on varactor, MEMS and MEMS capacitors. Tunable filters with half-cone transducer effectively suppress spurious modes, in contrast to other reported filters based on YIG thin films. Moreover, the frequency tuning range and bandwidth of this work are larger, and the out-of-band rejection is higher, especially for high frequency bands, when compared to other YIG thin film-based tunable passband filters.

## DISCUSSION

In this study, we present a novel half-cone transducer for YIG film-based MSW filters, achieving spurious-free and low-loss performance across an extremely wide frequency tuning range of up to 21.7 GHz. This transducer design is informed by Maxwell's equations and LLG equation combined with numerical simulations that clearly identify the mode shapes of the MSSW cavity filters. By tailoring spatial current distributions elaborately, the half-cone transducer can suppress generation of the width spurious modes and spatially couple more energy into the desired spin waves making up the passband of MSW cavity filter. Both numerical simulation and experiments validate that the half-cone transducer effectively suppresses spurious modes and reduces the insertion loss of single-cavity bandpass filters. Furthermore, a dual-cavity bandpass filter employing half-cone transducer is scaled to higher frequency bands with spurious-free performance. The insertion loss of 2.9–3.8 dB is lower than a co-fabricated reference straight-line bandpass filter. This study establishes a novel framework for tunable bandpass filter design, providing new insights into magnetostatic wave mode analysis and demonstrating the concurrent achievement of spurious modes suppression, broad frequency tuning, and low-loss performance through spatially tailored excitation of spin waves.

The spatially tailored spin wave technique in this work can be extended to other magnetostatic wave RF signal processing devices such as notch filters, delay lines, isolators, and circulators. The suppression of spurious modes in magnetostatic wave filters is indispensable for achieving high selectivity, minimal insertion loss, and reliable performance in advanced communication systems and sophisticated signal-processing applications.

## Methods

### MSWF Fabrication

The fabrication steps of YIG cavity filters follows the procedures in (*32*). The starting substrate was a 15 μm thick YIG thin film grown using liquid epitaxy on top of a 500 μm thick GGG substrate. GGG with <111> orientation has an excellent lattice coefficient and thermal expansion coefficient matching with YIG. First, a 500 nm $SiO_2$ film was deposited as a hard mask for YIG cavity etching using atomic layer deposition (ALD) using Bis (diethylamino) silane and trioxygen ($O_3$). After annealing the wafer at 600 °C in a nitrogen atmosphere, the $SiO_2$ layer was patterned using photolithography, development and dry etching with Trifluoromethane ($CHF_3$) and oxygen ($O_2$) gases. Then, the YIG layer was wet etched at around 150 °C in Phosphoric Acid. The etch rate is approximately 200 nm/min and the etch selectivity of YIG to $SiO_2$ was approximately 10:1. After etching the YIG layer, the remaining $SiO_2$ layer was stripped using 6:1 Buffered Oxide Etchant (BOE). For the 2 μm thick Al electrodes to be successfully laid on top of the 15 μm YIG cavities, a negative photoresist benzocyclobutene (BCB, CYCLOTENE 4026-46) was spin coated on the YIG and was patterned and developed using AP3000 developer, then was annealed at 300 °C for better stability and planarization. Next, an Al film was sputter deposited at 150 °C. Then, a 350 nm $SiO_2$ was deposited as a hard mask with 40 W RF power using Plasma Enhanced Chemical Vapor Deposition (PECVD). Next, the $SiO_2$ layer was patterned using photolithography and etched with $CHF_3$ and $O_2$ gases by reactive-ion etching (RIE). Finally, the Al layer was etched with boron trichloride ($BCl_3$) and chlorine ($Cl_2$) by inductively coupled plasma (ICP), and the remaining $SiO_2$ was removed by RIE etching.

### Measurement setup

The characterization of the YIG samples was carried out using a magnetic probe station (MicroXact, model MPS-

1D-5kOe). The required magnetic fields were produced via integrated electromagnets within the probe station. To ensure accuracy, the magnetic field strength was calibrated with a precision Gaussmeter. The frequency responses of the filters were obtained using a vector network analyzer (VNA, Keysight P5026B), operating under a standardized measurement condition with an input power level of -20 dBm and a consistent port impedance of 50 Ω. Before conducting the measurements, a Short-Open-Load-Thru (SOLT) calibration method was implemented to ensure accurate two-port calibration for the RF probes across the targeted frequency span. Ground-Signal-Ground (GSG) probes, featuring a probe gap of 150 μm, were employed to measure and extract the S-parameters of the MSSW filters. It should be noted that no additional on-wafer de-embedding corrections were applied in this study, meaning each filter was directly treated as the device under test (DUT). Further comprehensive details regarding the experimental setup and procedures are provided in Part 4 of the supplementary information.

## Supplementary Materials

**This PDF file includes:**

Supplementary Text

Figs. S1 to S10

Table S1

**Acknowledgments:** We would like to thank Dr. Todd Bauer, Dr. David Abe and Dr. Tim Hancock of the Defense Advanced Research Projects Agency (DARPA) and Dr. Michael Page of the Air Force Research Laboratory for their guidance and support of this work under the DARPA Wideband Adaptive RF Protection (WARP) program, contract FA8650-21-1-7010. **Funding:** The fabrication of devices was performed at the Singh Center for Nanotechnology, supported by the NSF National Nanotechnology Coordinated Infrastructure Program (No. NNCI-1542153). **Author contributions**: S.W. and S.Y. contributed equally to this work. S.W., S.Y. and R.O. developed the device concepts. S.W. and S.Y. built the simulation models of the devices. S.W. and X.D. fabricated the YIG filters. S.W., S.Y. and C.C. analyzed the measured data under the supervision of R.O. S.W., S.Y. and R.O. wrote and modified the manuscript. All authors have given approval to the final version of the manuscript. **Competing interests:** The authors declare that they have no competing interests**. Data and materials availability**: All data supporting the findings of this study are available within the article and the supplementary materials. Any additional requests for information can be directed to and will be handled by the corresponding author.

Supplementary Materials for

# Spatially tailored spin wave excitation for spurious-free, low-loss magnetostatic wave filters with ultra-wide frequency tunability

Shuxian Wu, Shun Yao *et al.*
*Corresponding author. Email: rolsson@seas.upenn.edu

**This PDF file includes:**

## 1. Dispersion of MSSW

Magnetostatic wave (MSW) propagation modes are divided into three categories, including Magnetostatic Forward Volume Waves (MSFVW), Magnetostatic Backward Volume Waves (MSBVW), and Magnetostatic Surface Waves (MSSW) (*47*). MSFVWs are generated when the bias magnetic field $\vec{H}_0$ is applied perpendicular to the surface of the ferrite slabs such that $\vec{H}_0 \cdot \vec{n} = 0$. The wave vector ($\vec{k}$) is perpendicular to the applied magnetic field. MSFVWs are characterized by symmetric propagation throughout the volume of the slabs, and are often preferred in systems requiring minimal dispersion with high phase stability. MSBVWs appear when the bias magnetic field is parallel to $\vec{k}$, and both lie within the plane of the ferrite slabs. Such implementation results in backward wave behavior, where the phase velocity and group velocity are in opposite directions. The frequency dispersion of MSBVWs is negatively correlated to the wavenumber and is typically used in delay line and signal processing applications. MSSWs occur when $\vec{H}_0$ is perpendicular to $\vec{k}$, i.e., the bias field lies in the plane of the film and is perpendicular to the direction of propagation. This configuration gives rise to highly localized surface-confined waves that exhibit non-reciprocal behavior and strong field confinement near the surface of the ferrite film, making them particularly advantageous in non-reciprocal devices such as filters and isolators. Therefore, we focus on MSSW filters in this study.

Figure S1 shows the dispersion curve of MSSWs in a 15 μm thick, thin-film YIG with a width of 400 μm and magnetic field intensity of 2500 Gauss. Different colors represent different orders of width modes ($m$ = 1, 2, 3,…), and $m$ = 0 indicates the MSSW propagation modes when $k_y$ is equal to 0, representing a thin film YIG with infinite width. When the MSSW wavelength is comparable to or greater than the width of YIG, the width of the sample has an effect on the dispersion. The dispersion curve shifts toward lower frequencies as the order of the width mode increases. This phenomenon indicates the strong potential for spurious modes directly below the passband (lower frequency) for MSSW filters. A reference frequency ($f_1$) is the resonant frequency corresponding to the first order primary mode with a first order width mode. As $k_x$ increases, the order of the primary mode increases and the frequency interval of different width modes becomes smaller. The dispersion analysis of the MSSW here provides a theoretical guide for the later simulations and experiments.

## 2. Current density distribution

For the magnetostatic surface wave (MSSW) cavity filters, the distribution of current density varies among transducers with different widths or volumes. To study the variation of the current density in MSSW cavity filters with full-cone and half-cone transducers, simulations are implemented using the Ansys HFSS platform. For the reference straight-line bandpass cavity filter without apodization, the current density is high across the entire length of the transducers, as illustrated in Figure S2A. When the full-cone and half-cone transducers are adopted, the current density distributions are depicted in Figure S1B-S1C, which include both high current density and low current density regions due to the volume change of the aluminum (Al) electrodes.

Specifically, the current densities in the full-cone and half-cone regions with wide electrodes are much lower than those in the central area with narrow electrodes. Moreover, for both full-cone and half-cone configurations, the radio-frequency (RF) current is predominantly concentrated at the edges, with extremely low current density in the interior region. These observations indicate that the apodization techniques effectively modify the current density distribution in the bandpass filters, and the transducer structures can be designed to obtain the desired primary modes (w = 1) and suppress spurious modes (w > 1). This behavior influences the dynamic magnetization of the YIG film as well as the excitation of width modes, providing a theoretical guidance for the spurious modes suppression.

## 3. Optimization of half-cone apodization

To achieve optimal spurious suppression for the half-cone bandpass filters, the dimensions of the half-cone transducer structure were optimized at a magnetic field intensity of 7500 Gauss. Figure S3 shows the top view of the three-dimensional (3D) half-cone bandpass filter model in simulations, where HC_X represents the length of the half-cone along the *x*-direction, and HC_Y represents the length of the half-cone along the *y*-direction. The curved shape of the half-cone is defined by the three-point arc method in Ansys HFSS. Initially, the bandpass filter without apodization is simulated, and there are some spurious modes in the frequency response as shown in Figure S4A. Next, the HC_X was optimized. Figure S4B-S4H presents the simulated frequency responses of filters with half-cone transducer apodization with different dimensions for HC_X. Specifically, HC_X was varied from 40 μm to 70 μm in increments of 5 μm, while the HC_Y was held constant at 100 μm. Through comparative analysis, the half-cone filter with a HC_X of 65 μm can effectively suppress spurious modes and provide a flatter passband response.

Subsequently, the HC_Y dimensions were optimized by scaling the size from 60 μm to 140 μm in increments of 20 μm, while maintaining HC_X at the optimized length of 65 μm, as shown in Figure S5. It is seen that the half-cone filter when HC_Y is 100 μm achieved high spurious mode suppression and excellent passband flatness simultaneously. The simulation optimizations here serve as a valuable guideline for subsequent experiments.

## 4. Measurement setup

Figure S6 shows the measurement setup utilized to measure the frequency response of the MSSW filters. Figure S6A illustrates the overall measurement configuration, comprising a vector network analyzer (Keysight, P5026B) for transmitting and receiving RF signals, a magnetic probe station (MicroXact's MPS-1D-5kOe) connected with a Gaussmeter (Model GM2, AlphaLab Inc) for precise manipulation of the external bias magnetic fields, and an RF coaxial cable for signal transmission. Additionally, a microscope is employed to enable accurate positioning and observation, whereas the probe carrier with vacuum pump system facilitates stable contact between the probes and the filters under test. Electromagnets and a magnetic field controller are used to produce the external bias magnetic fields. Figure S6B provides a detailed top view of the testing area, clearly showing the sample with several MSSW filters positioned centrally between two electromagnets. Electromagnets are strategically placed on either side of the filters to generate the bias magnetic field, which is critical for testing the frequency response and tunability of the filters.

## 5. Full-cone, extended-cone and half-cone transducers comparison

Figure S7 provides scanning electron microscopy (SEM) images of the filters employing two different transducer apodization designs: a full-cone structure (Fig. S7A) and an extended-cone structure (Fig. S7B). The full-cone transducer apodization is characterized by symmetric electrode shapes, as indicated by the red double arrows representing uniform transducer width. In contrast, the extended-cone transducer features an asymmetric electrode geometry resulting in different widths which are marked by the red label "shorter" and yellow label "longer". The structural design influences the electromagnetic coupling behavior within the filter, potentially affecting spurious mode suppression and impedance matching properties. The SEM images illustrate the geometric differences essential for achieving distinct performance characteristics in magnetostatic wave filters.

The measured $S_{12}$ frequency responses of the filters employing full-cone and extended-cone transducer apodization at different magnetic flux density intensities ($B$) are depicted in Figure S8, which illustrate the effectiveness of spurious mode suppression and insertion loss (IL) characteristics. Specifically, Figure S8 shows the responses of the reference filters (shown in blue), the filters with full-cone apodization (purple), and the filters with extended- cone apodization (red). The introduction of both full-cone and extended-cone transducer apodization notably suppresses spurious responses compared to the reference straight-line filter, as indicated by smoother passband transitions and reduced out-of-band spurious peaks. Notably, the full-cone apodization provides moderate spurious suppression since there still exits a spurious mode on the left of the passband, whereas the extended-cone structure achieves better spurious suppression effect. Although full-cone and extended-cone transducer apodization can suppress spurious modes, these do so at the expense of filter insertion loss and bandwidth. For the extended-cone filters, the insertion loss is much larger and the bandwidth is smaller compared to the reference filters and full-cone filters. It can be concluded that the extended-cone apodization suppresses spurious modes well, but degrades the passband performance and increases the insertion loss of the filters.

To further compare the performance of full-cone and half cone transducer apodization, Figure S9 presents the measured $S_{12}$ frequency responses of the reference straight-line bandpass filters, along with those employing full-cone and half-cone transducer apodization under different bias magnetic field intensities. A comparative analysis reveals that the filter using the half-cone apodization design exhibits superior overall performance compared to the full-cone apodization design. Specifically, the half-cone transducer effectively mitigates spurious mode excitation, achieves lower insertion loss, and yields a flatter passband profile. These results indicate that the half-cone apodization approach offers the most efficient suppression of spurious modes while improving accuracy and selectivity in MSSW bandpass filter applications.

## 6. Smith chart of dual-cavity bandpass filters

Figure S10 presents the Smith chart diagrams of the measured $S_{11}$ parameters for the reference and half-cone dual-cavity bandpass filters in the frequency range of 18 GHz to 32 GHz under various magnetic field intensities. The blue curves correspond to the reference filter, while the red curves indicate filters with the half-cone transducers. A notable difference is evident between the two filters, primarily attributed to the altered transducer volume resulting from the half-cone geometry.

Utilizing the Smith charts depicted in Figure S10, the resonant frequency corresponding to the minimum insertion losses ($f_{IL\min}$) were identified. The yellow markers denote the position of $f_{IL\min}$ for the reference filters, whereas the green markers indicate the position of $f_{IL\min}$ for the half-cone filters. The corresponding complex impedances are listed in the lower left corner of the circle chart. It is clearly observable across all frequency bands that the half-cone design exhibits superior impedance matching compared to the reference filters, as demonstrated by their positions closer to the center of the Smith chart. Specifically, within the frequency range of 18–25 GHz (corresponding to magnetic field intensities of 5500–7500 Gauss), the half-cone filters in this work achieve optimal impedance matching and the lowest insertion losses. The results further validate that the half-cone transducer effectively enhances overall performance of the MSSW filters by spatially exciting only the cavity regions where the magnetostatic waves comprising the filter passband are located.

## 7. Performance comparison of tunable bandpass filters

Table S1 provides a performance comparison between the tunable bandpass filters reported in this study and those based on varactors, MEMS, and previously reported thin film YIG-based technologies *(11), (14), (28), (31), (53–58)*. While varactor and MEMS-based filters offer advantages in integration, they often suffer from limited tuning range, high insertion loss, limited rejection, or large footprint. For instance, MEMS filters exhibit wide bandwidths and strong rejection but typically occupy areas of over 100 mm². YIG-based filters are known for their low loss and high spectral selectivity. However, the simultaneous realization of spurious modes suppression and maintaining low insertion loss over a wide frequency tuning range has not been previously demonstrated.

The tunable filters reported in this study demonstrate a wide frequency tuning range, low insertion loss and compact size. The single-cavity design covers 6.3 – 16.8 GHz with a 2.7:1 tuning ratio, and an insertion loss as low as 2.4 dB, with a footprint of 0.55 mm². The dual-cavity variant further expands the tuning range to 9.8 – 31.5 GHz, achieving a 3.2:1 tuning ratio while maintaining low loss (2.9 – 3.8 dB) and high out-of-band suppression up to 41 dB, in a footprint of just 1.09 mm². The half-cone bandpass filters in this work present spurious-free responses and low insertion loss over an unprecedented frequency tuning range of 21.7 GHz. The comparison results position the reported bandpass filters among the most compact and efficient YIG thin film-based solutions, highlighting their potential for integration in future reconfigurable wireless communication systems.

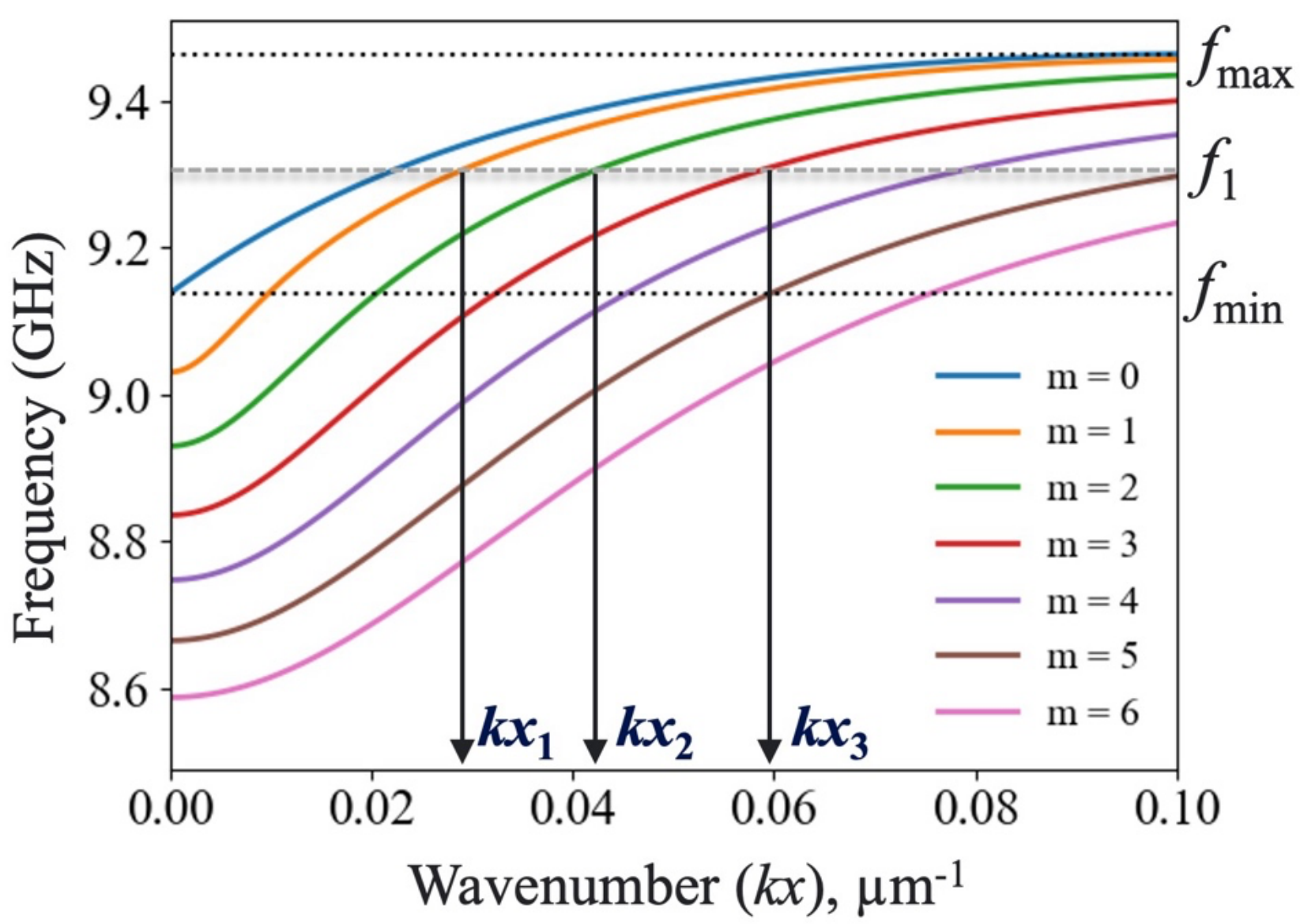

**Fig. S1.** Dispersion of YIG thin-film cavity based MSSW propagation. Different colors represent different orders of width modes. $f_{max}$ and $f_{min}$ represent the maximum frequency and minimum frequency, and $f_1$ is the resonant frequency corresponding to the first-order primary mode with a first-order width mode. As $k_x$ increases, the order of the primary modes increases and the frequency interval of different order width modes becomes smaller.

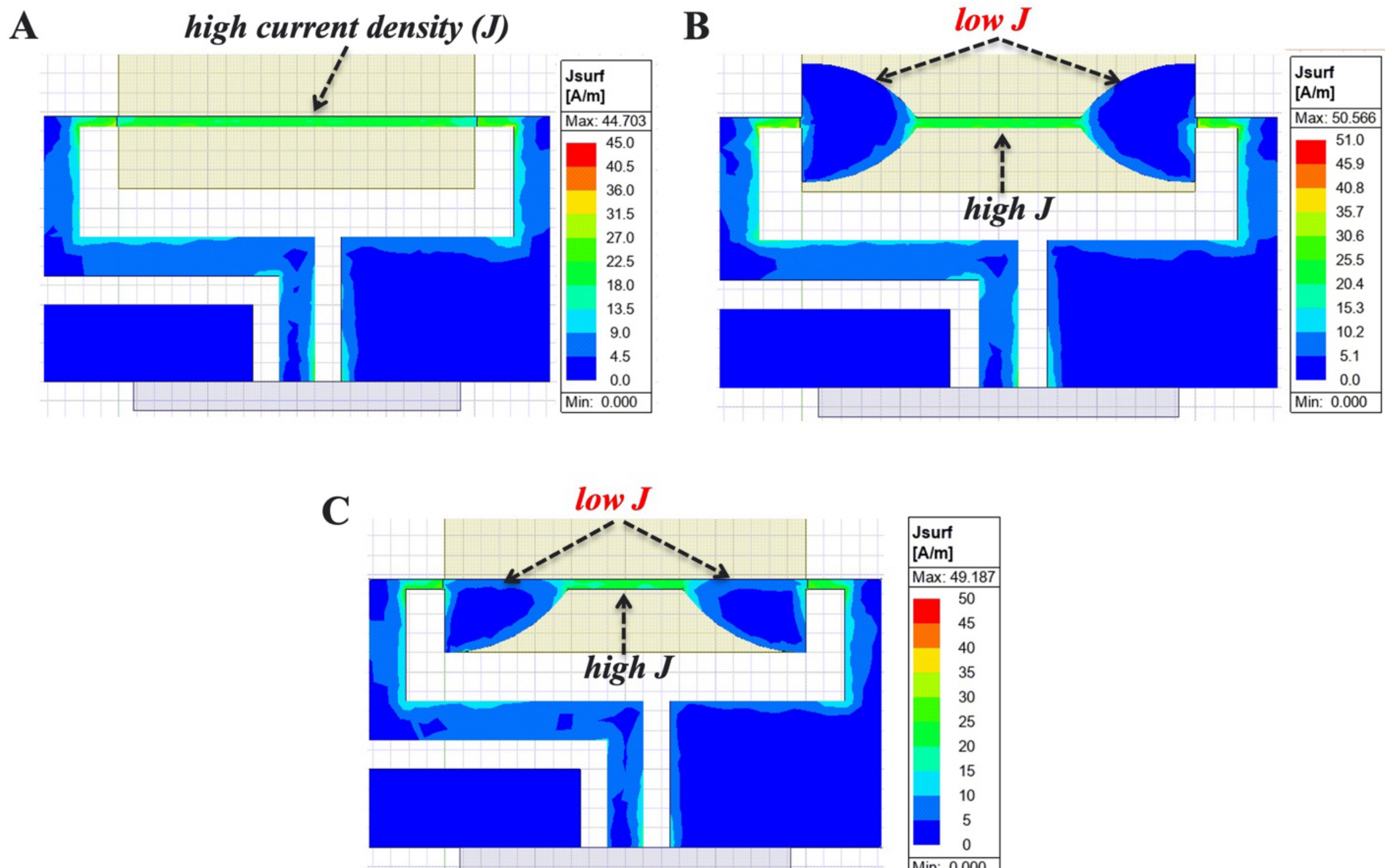

**Fig. S2. The simulated current density distributions for the three MSSW bandpass cavity filters.** (**A**) Reference straight-line bandpass cavity filter without apodization. (**B**) Bandpass filter with full-cone transducers. (**C**) Bandpass filter with half-cone transducer. After adopting the full-cone and half-cone transducer designs, the current density distributions are spatially tailored to suppress spurious modes (w>1), while the half-cone transducer is further tailored to efficiently excite the primary modes (w=1).

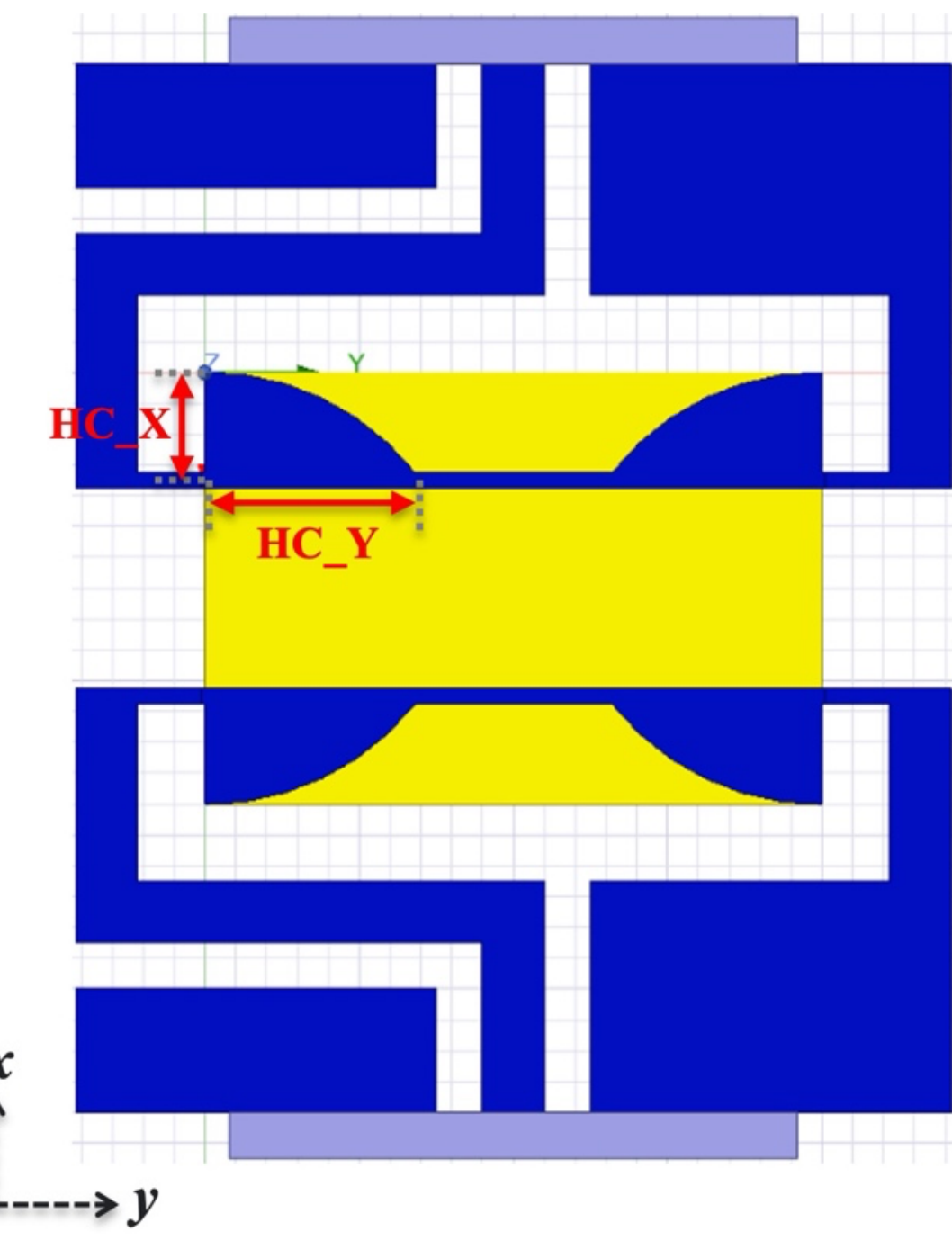

**Fig. S3.** Top view of the three-dimensional (3D) half-cone bandpass cavity filter model in simulations, where HC_X represents the size of the half-cone transducer along the *x*-direction, and HC_Y represents the size of the half-cone transducer along the *y*-direction. The shape of the half-cone transducer is defined by the three-point arc method.

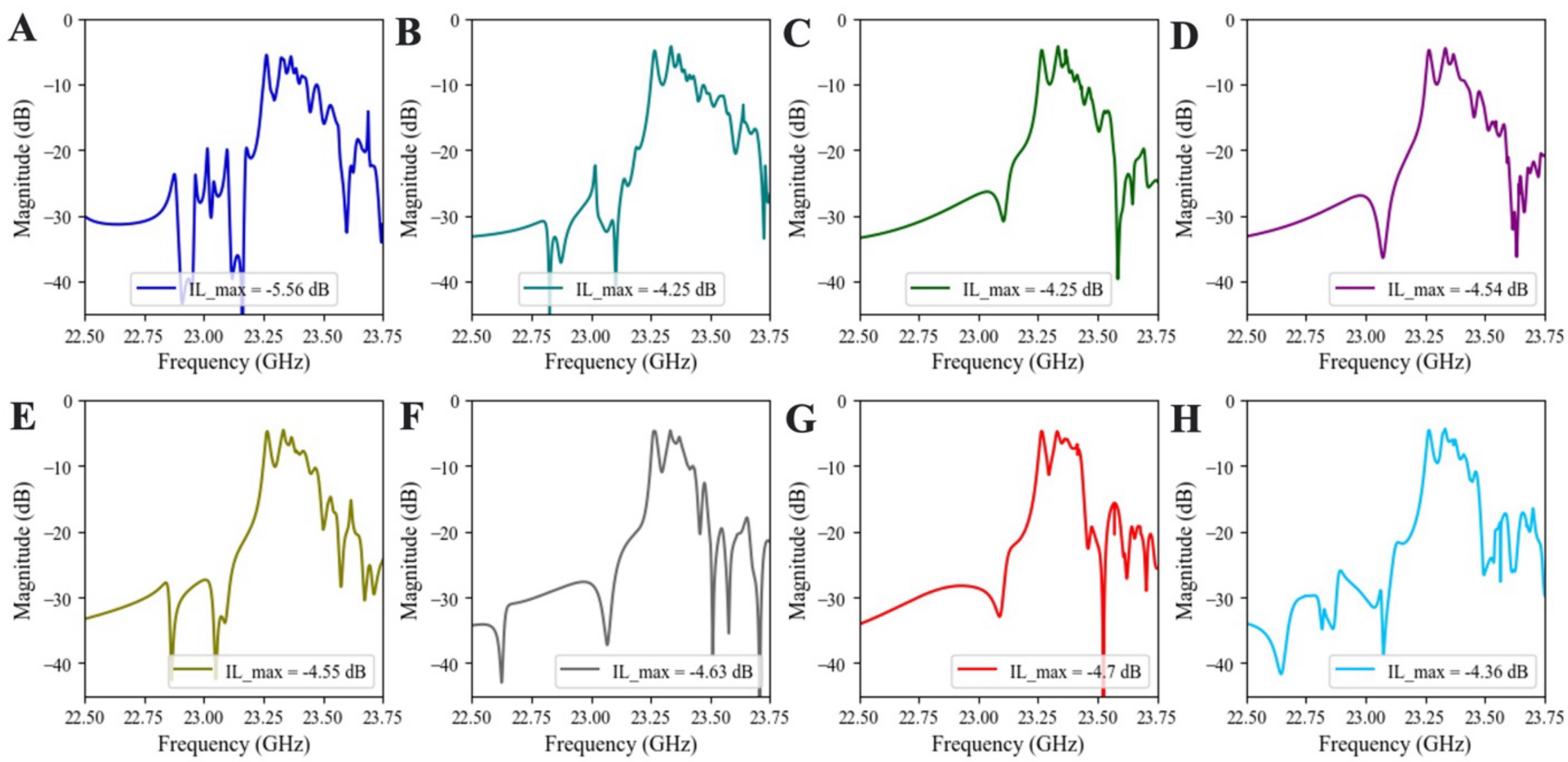

**Fig. S4. Frequency responses of reference straight-line and half-cone bandpass cavity filters with different lengths along the *x*-direction (HC_X) when the length along the *y*-direction (HC_Y) is 100 μm.** (**A**) Without half-cone transducer apodization. (**B - H**) HC_X = 40 μm, 45 μm, 50 μm, 55 μm, 60 μm, 65 μm, and 70 μm. It is observed that the half-cone transducer apodization with a HC_X of 65 μm effectively suppresses spurious modes and provides a flatter passband response.

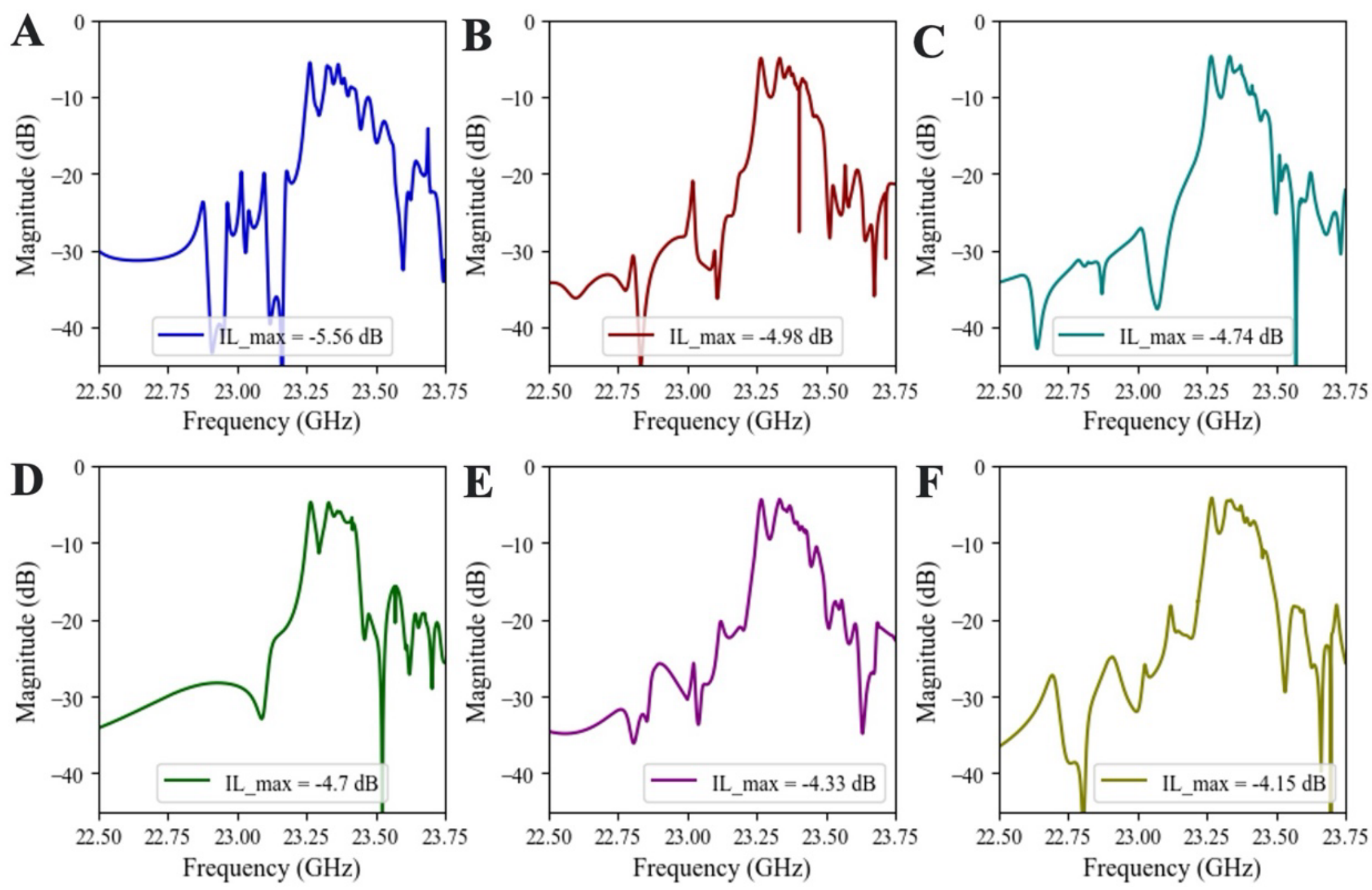

**Fig. S5. Frequency responses of reference straight-line and half-cone bandpass cavity filters with different HC_Y when HC_X is 65 µm.** (**A**) Without half-cone apodization. (**B - F**) HC_Y = 60 µm, 80 µm, 100 µm, 120 µm, and 140 µm. It is seen that the half-cone cavity filter when HC_Y is 100 µm achieved strong spurious mode suppression and excellent passband flatness simultaneously.

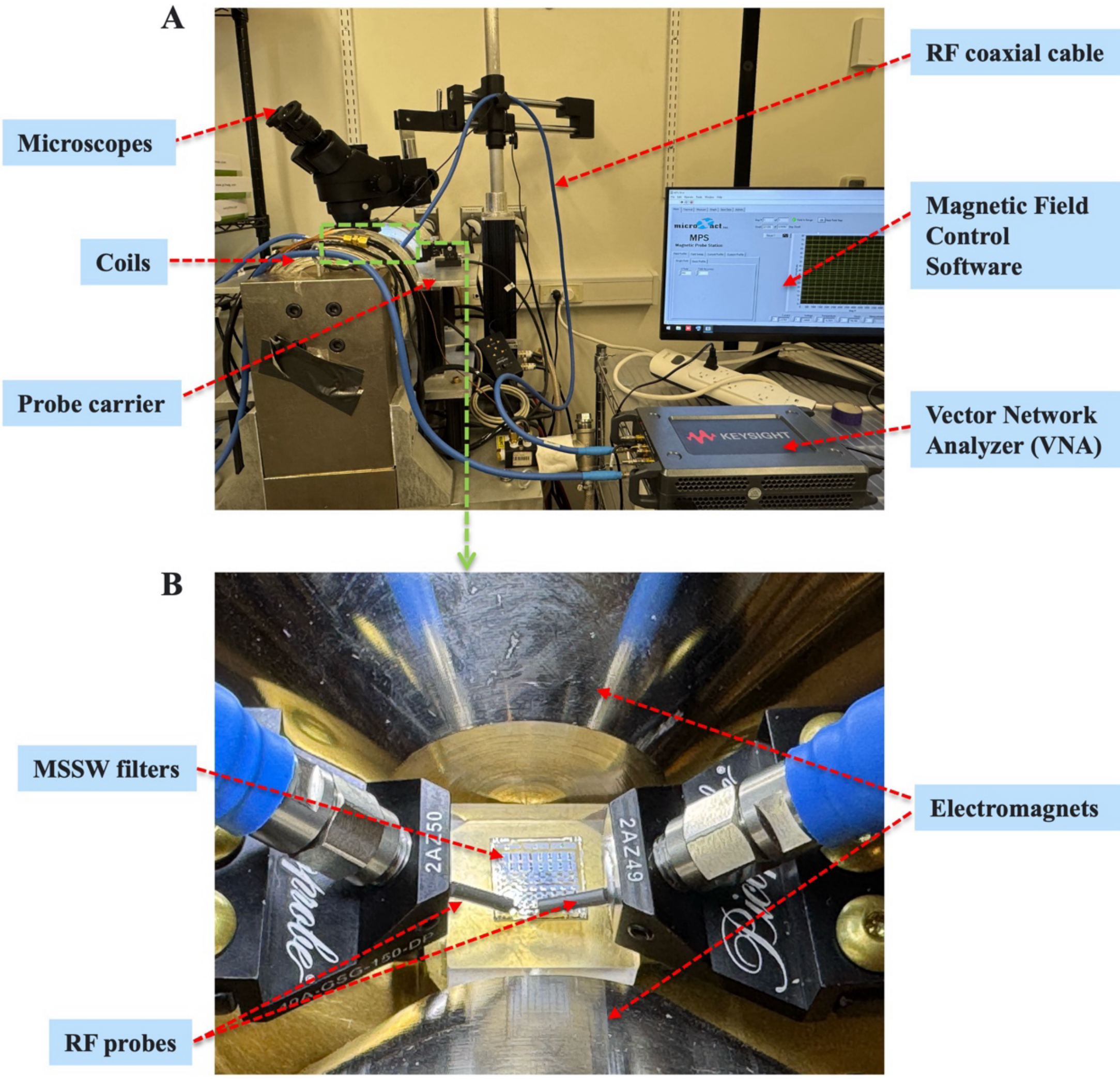

**Fig. S6. Measurement setup utilized to measure the frequency response of the MSSW filters. (A)** Overall measurement configuration. (**B)** Top view of the testing area which shows the sample with several MSSW filters positioned centrally between two electromagnets.

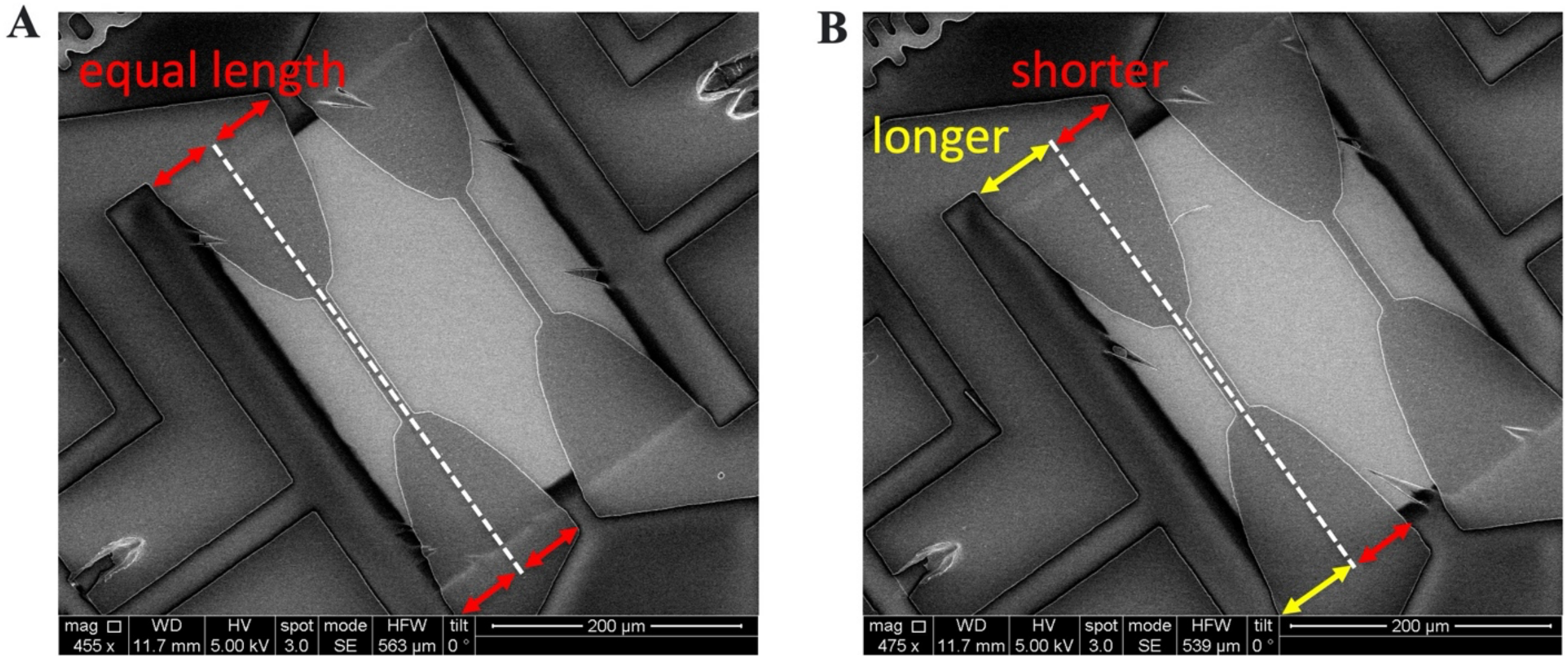

**Fig. S7. The scanning electron microscopy (SEM) images of the fabricated filters employing two different transducer designs.** (**A)** Full-cone transducer apodization bandpass cavity filter. (**B)** Extended-cone transducer apodization bandpass cavity filter.

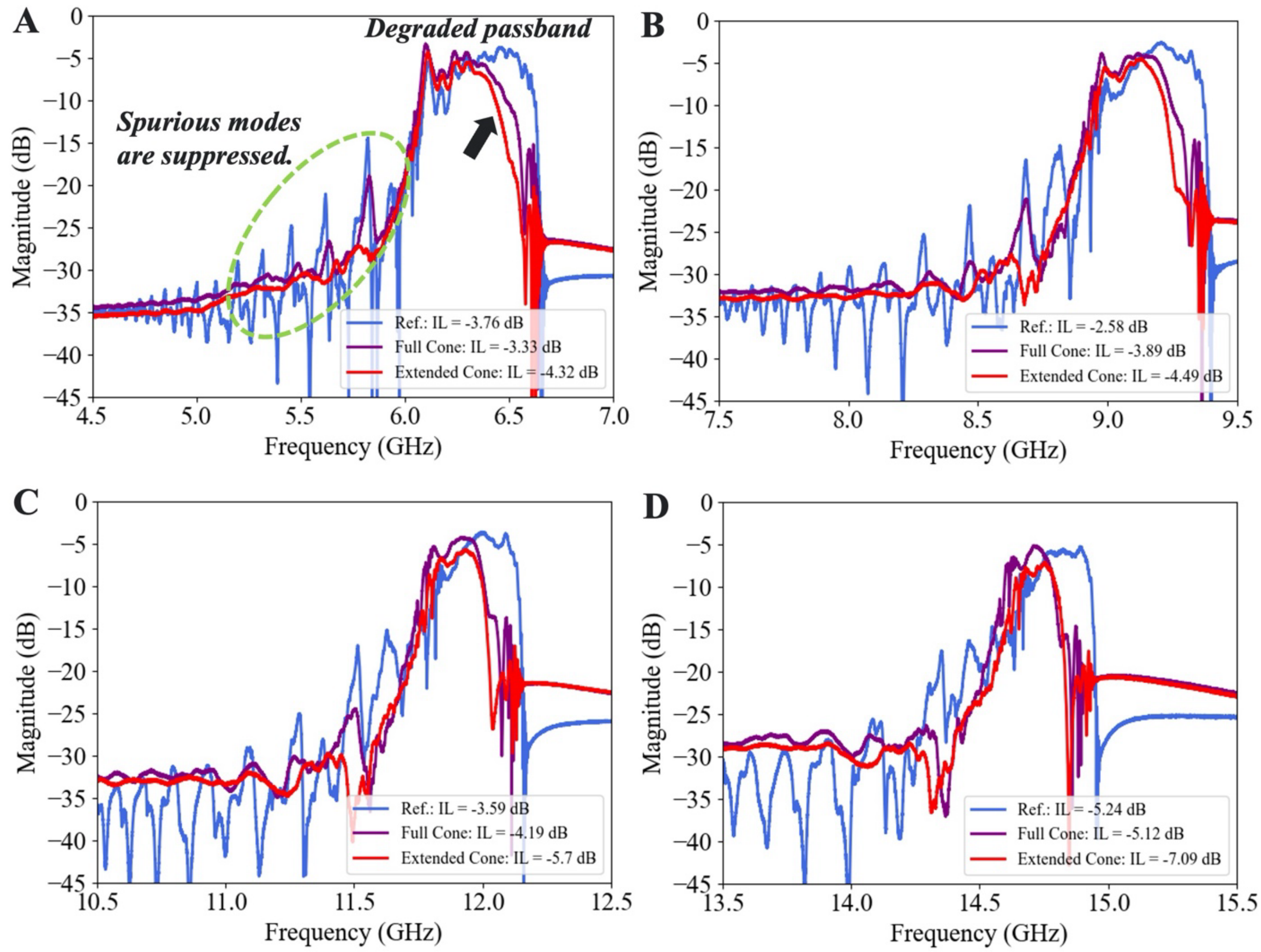

**Fig. S8. Measured $S_{12}$ frequency responses of the reference straight-line cavity filter and filters employing full-cone and extended-cone transducer apodization designs at different magnetic flux density intensities (*B*).** (**A-D**) $B$ = 1500 Gauss, 2500 Gauss, 3500 Gauss, and 4500 Gauss. Both the full-cone and extended-cone transducers suppress spurious responses compared to the reference straight-line structure. The full-cone transducer provides moderate spurious suppression effect since there still exits a spurious mode on the left of the passband, whereas the extended-cone structure achieves better spurious suppression performance. However, the extended-cone structure degrades the passband performance and increases the insertion loss of the cavity filters.

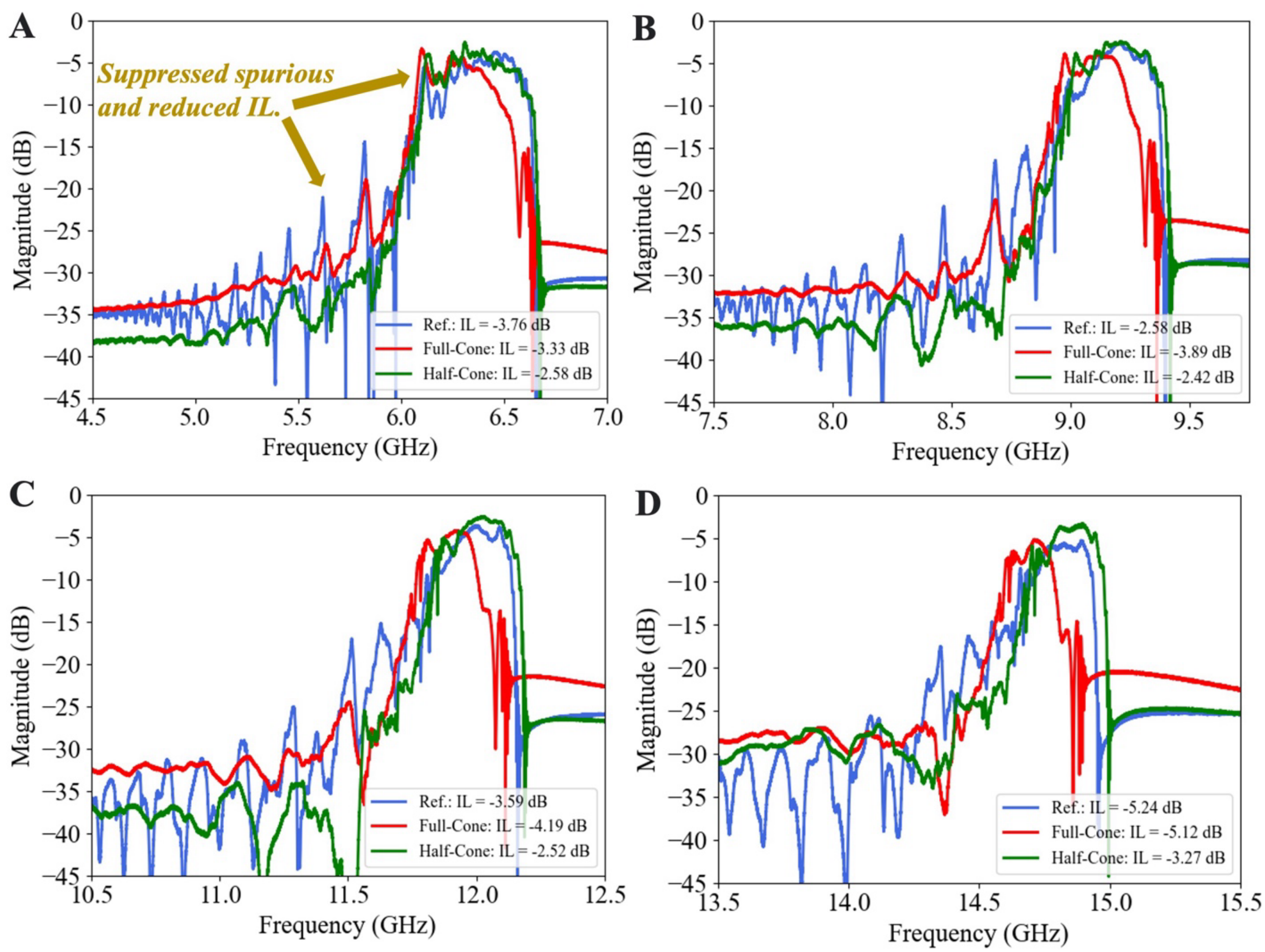

**Fig. S9. Measured $S_{12}$ frequency responses of the reference straight-line filter and filters employing full-cone and half-cone transducer apodization designs at different magnetic bias fields**. (**A-D)** $B$ = 1500 Gauss, 2500 Gauss, 3500 Gauss, and 4500 Gauss. It can be clearly observed that the overall performance of the filter with half-cone transducer apodization is better than that with the full-cone transducer. The half-cone transducer apodization not only effectively suppresses spurious modes, but also reduces insertion loss and makes the passband flatter.

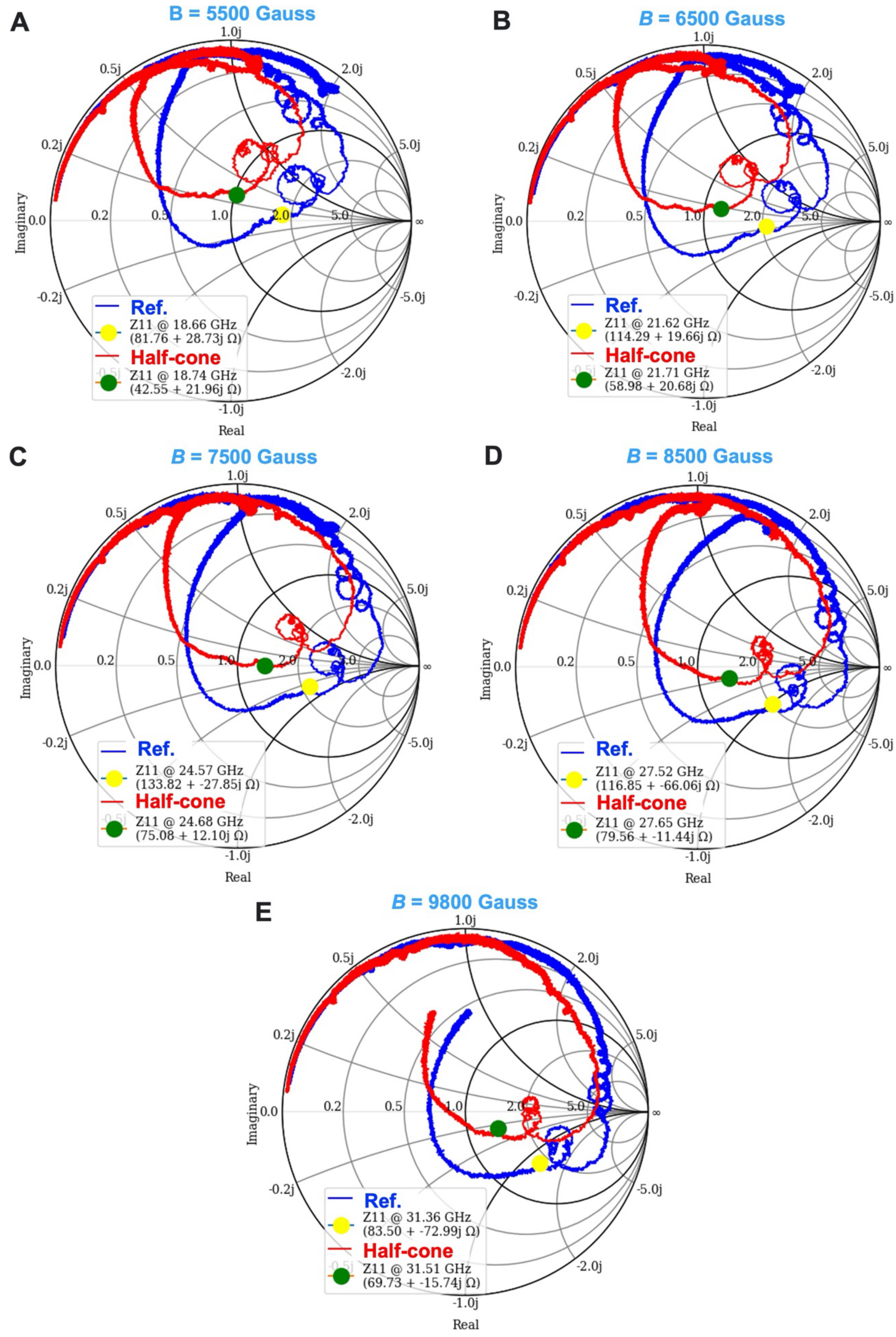

**Fig. S10. Smith chart diagrams of the measured $S_{11}$ parameters for the reference and half-cone dual-cavity bandpass filters in the frequency range of 18 GHz to 32 GHz under various magnetic field**

**intensities.** (**A-E**) $B$ = 5500 Gauss, 6500 Gauss, 7500 Gauss, 8500 Gauss, and 9800 Gauss. It is observed across all frequency bands that the half-cone transducer design exhibits superior impedance matching compared to the reference straight-line transducer filters due to their positions closer to the center of the Smith charts.

| Reference | Frequency Tuning (GHz) | Tuning Ratio | Insertion Loss (dB) | Rejection (dB) | Bandwidth (MHz) | Area ($mm^2$) |
|---|---|---|---|---|---|---|
| Varactor [11] | 10.23–15.7 | 1.5:1 | 2.6–3.9 | 24–39 | 5,120–11,540 | 743 |
| Varactor [53] | 0.6–1.02 | 1.7:1 | 1.1–2.8 | 26–31 | 93–157 | 9025 |
| Varactor [54] | 1.7–2.1 | 1.2:1 | 2.84–2.9 | N/A | 85–105 | 783 |
| MEMS [14] | 6.5–10 | 1.5:1 | 4.1–5.6 | N/A | 306–539 | 20 |
| MEMS [55] | 23.4–35.1 | 1.5:1 | 1.5–4.2 | 33–39 | 200–1,400 | 112 |
| MEMS Capacitors [56] | 18.6–21.4 | 1.2:1 | 3.85–4.15 | N/A | 1,395–1,605 | N/A |
| YIG [28] | 4.5–10.1 | 2.2:1 | 3.55–6.94 | 25–35 | 11–39 | 0.64 |
| YIG [31] | 3.4–11.1 | 3.3:1 | 3.2–5.1 | 23.5–35 | 18–25 | 60 |
| YIG [57] | 5–19 | 3.8:1 | 1.7–2 | 20–25 | 62–107 | 70 |
| YIG [58] | 3.3–9.7 | 2.9:1 | 2.1–3 | 35–45 | 68–264 | 1.03 |
| **This study (Single-Cavity)** | 6.3–16.8 | 2.7:1 | 2.4–3.2 | 25–39 | 181–211 | 0.55 |
| **This study (Dual-Cavity)** | 9.8–31.5 | 3.2:1 | 2.9–3.8 | 23–41 | 59–201 | 1.09 |

**Table S1.** Performance comparison with other tunable bandpass filters.